\documentclass[pre,twocolumn,showpacs,floatfix]{revtex4}
 \usepackage{graphicx}
 \usepackage{epstopdf}
 \usepackage{dcolumn}
 \usepackage{color}
 \usepackage{bm}
 \usepackage{verbatim}
 \usepackage{multirow}
 \usepackage{amsmath,amssymb,graphicx}
 \usepackage{epsfig}
 \usepackage{enumerate}
 \usepackage{array}
 \usepackage{multirow}
\newcommand{\figone}
{
	\begin{figure*}[h]
		\begin{center}
		\includegraphics[width=6.0in,height=2.5in]{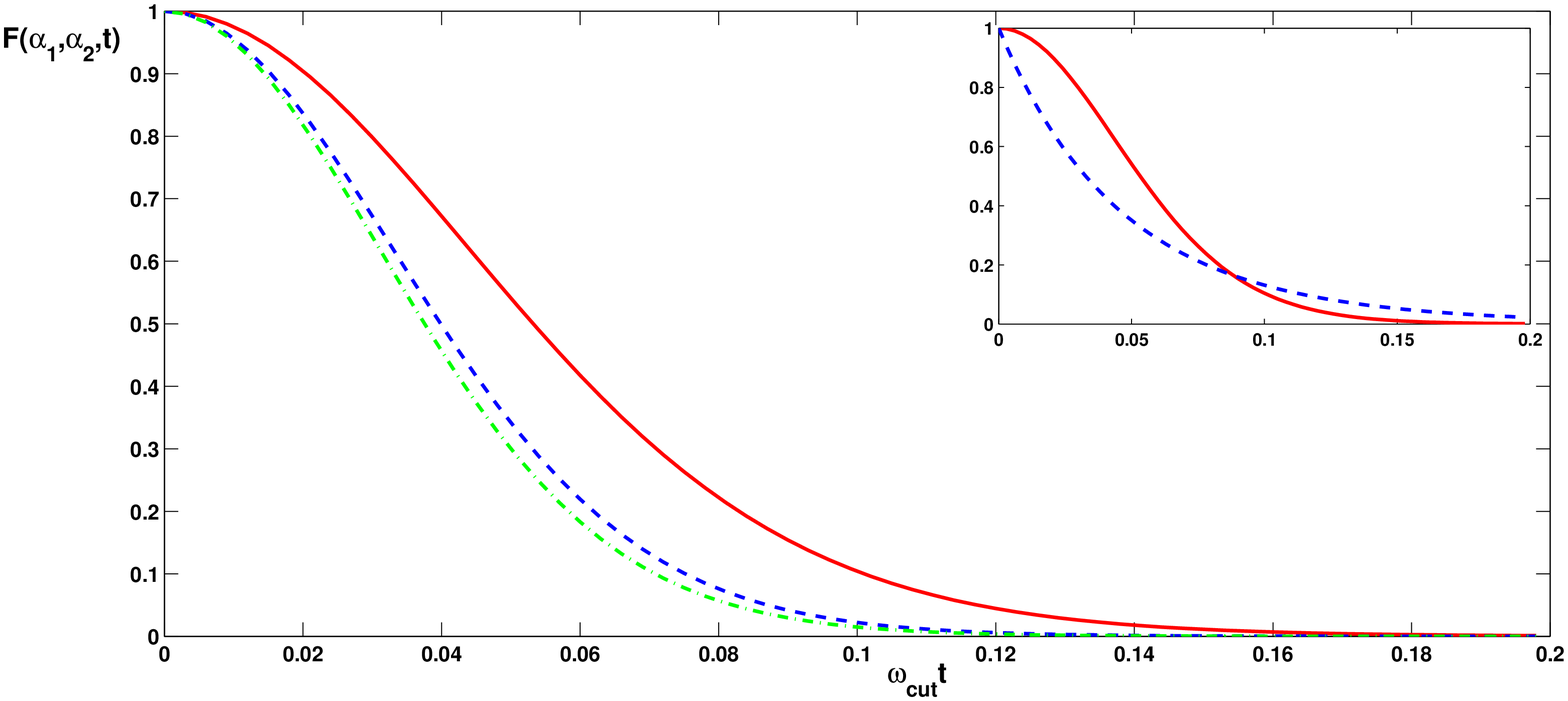}
		\caption{ The dynamics of the fringe visibility are plotted as a function of time $t$ in arbitrary unit  for the Ohmic (red solid line), sub-Ohmic (blue dashed line) and super-Ohmic (green dashed dotted line) reservoirs at dimensionless temperature $k_BT/\hbar\omega_{1,2}=100$. The inset shows the comparison  between the non-Markovian (red solid line) and the Markovian (blue dotted line) fringe visibility for the Ohmic reservoir. We use $\omega_{cut}=1$, $r_0=6.0$ and $r_c=4.0$ for the $r_{1,2}>>1$ regime. }
		\end{center}
	\end{figure*}
}

\newcommand{\figtwo}
{
	\begin{figure*}[h]
		\begin{center}
		\includegraphics[width=6.0in,height=2.5in]{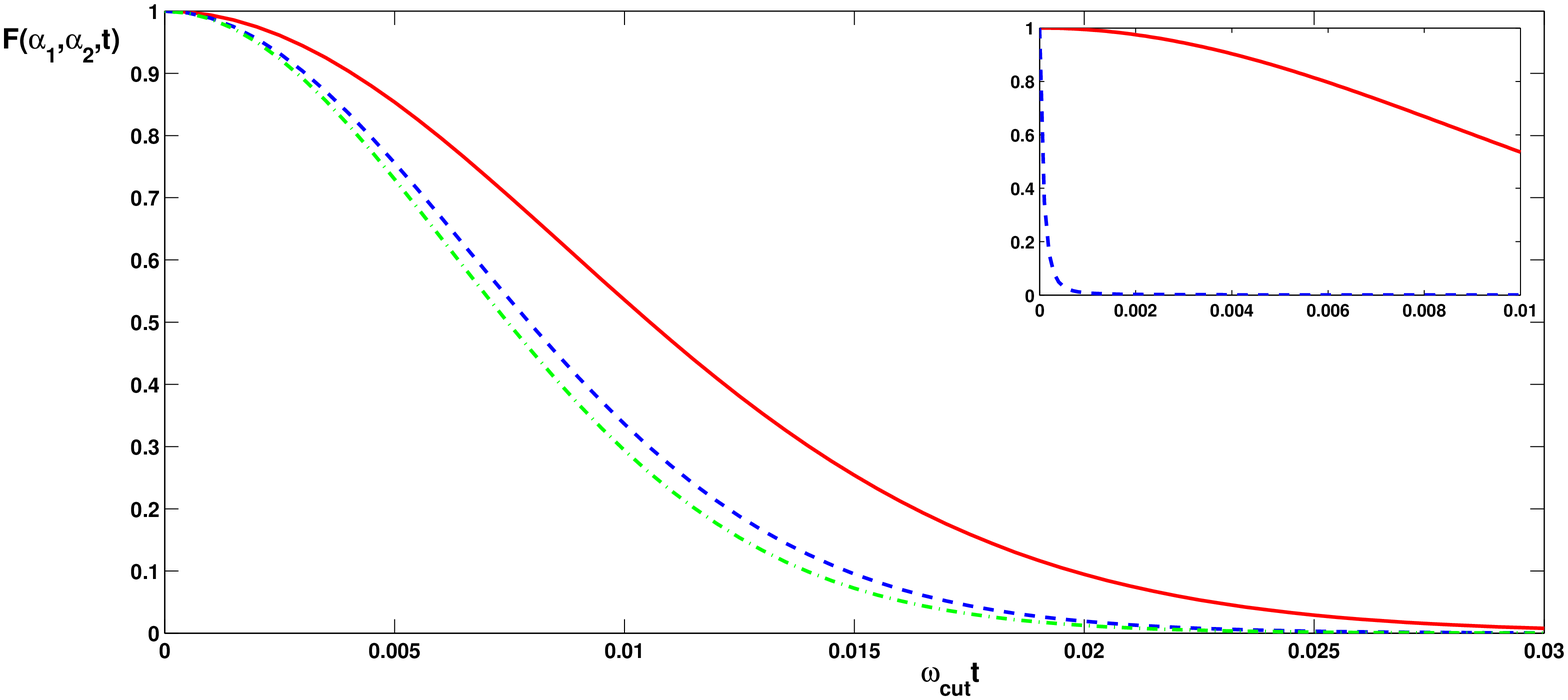}
		\caption{ We plot the fringe visibility  as a function of time $t$ (in arbitrary unit) for the Ohmic (red solid line), sub-Ohmic (blue dashed line) and super-Ohmic (green dashed dotted line) reservoirs at a dimensionless temperature $k_BT/\hbar\omega_{1,2}=100$. In the inset, we plot the fringe visibility for the non-Markovian (red solid line) and the Markovian (blue dotted line) cases for the Ohmic reservoir.  We use $\omega_{cut}=1$, $r_0=0.5$ and $r_c=0.05$ for the $r_{1,2}<<1$ regime. }
		\end{center}
	\end{figure*}
}
\newcommand{\figthree}
{
	\begin{figure*}[h]
		\begin{center}
		\includegraphics[width=7.0in,height=3.5in]{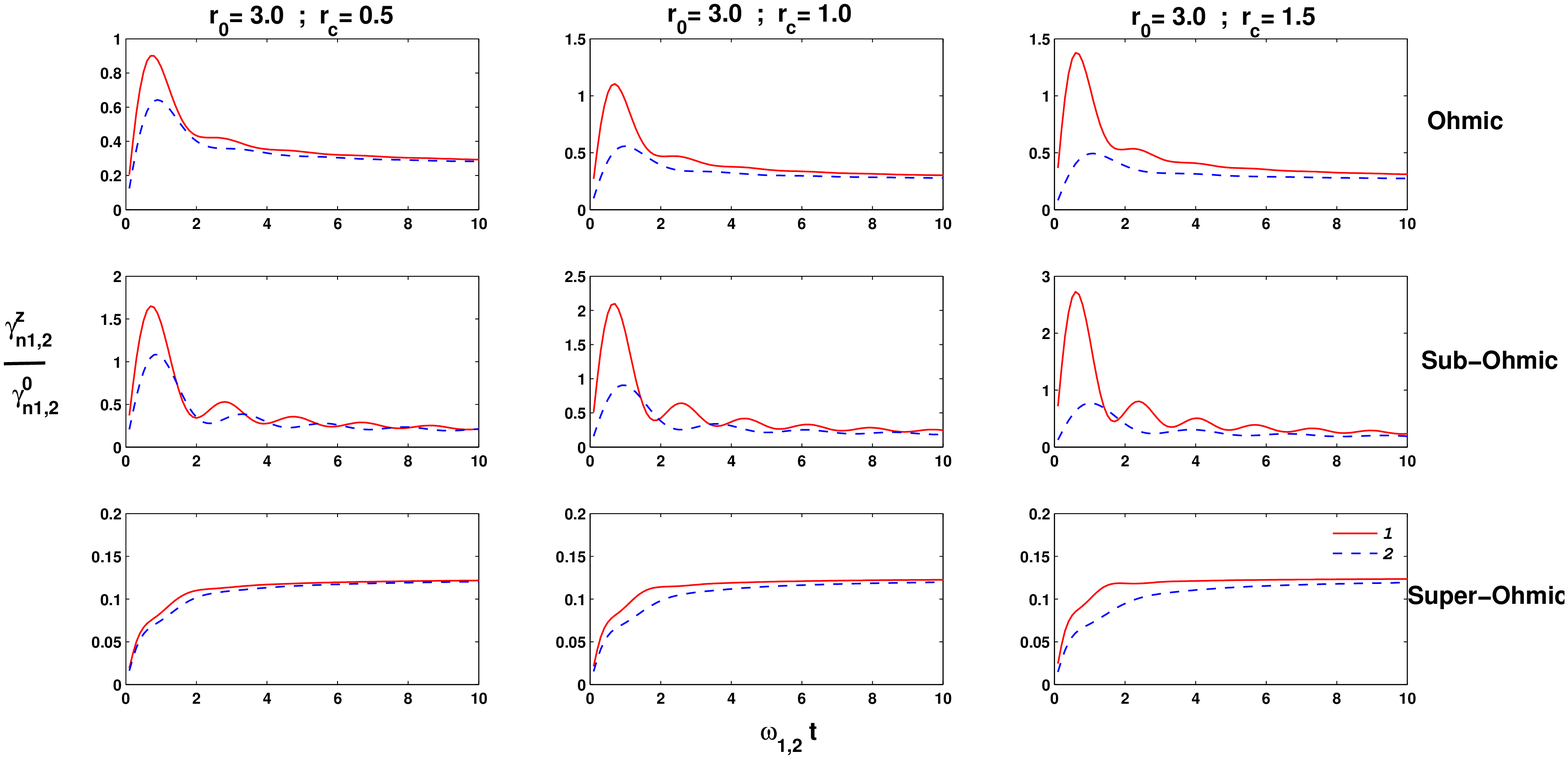}
		\caption{ Dynamics of the ratio between the effective decay rate $\gamma_{n_{1,2}}^z(\tau)$ and the Markovian decay rate $\gamma_{n_{1,2}}^0$ are plotted as a function of unitless time $\omega_{1,2}t$  for the Ohmic (uppermost row), sub-Ohmic (middle row) and super-Ohmic (lowermost row) reservoirs at high temperatures for a fixed value of $r_0=3.0$ and a variation of magnetic field parameter $r_c$. We use $\omega_{cut}=1$ for the $r_{1,2}>1$ regime, where the three columns denote $r_c=0.5$ (left), $r_c=1.0$ (middle), and $r_c=1.5$ (right) cases, respectively. }
		\end{center}
	\end{figure*}
}

\newcommand{\figfour}
{
	\begin{figure*}[h]
		\begin{center}
		\includegraphics[width=7.0in,height=3.5in]{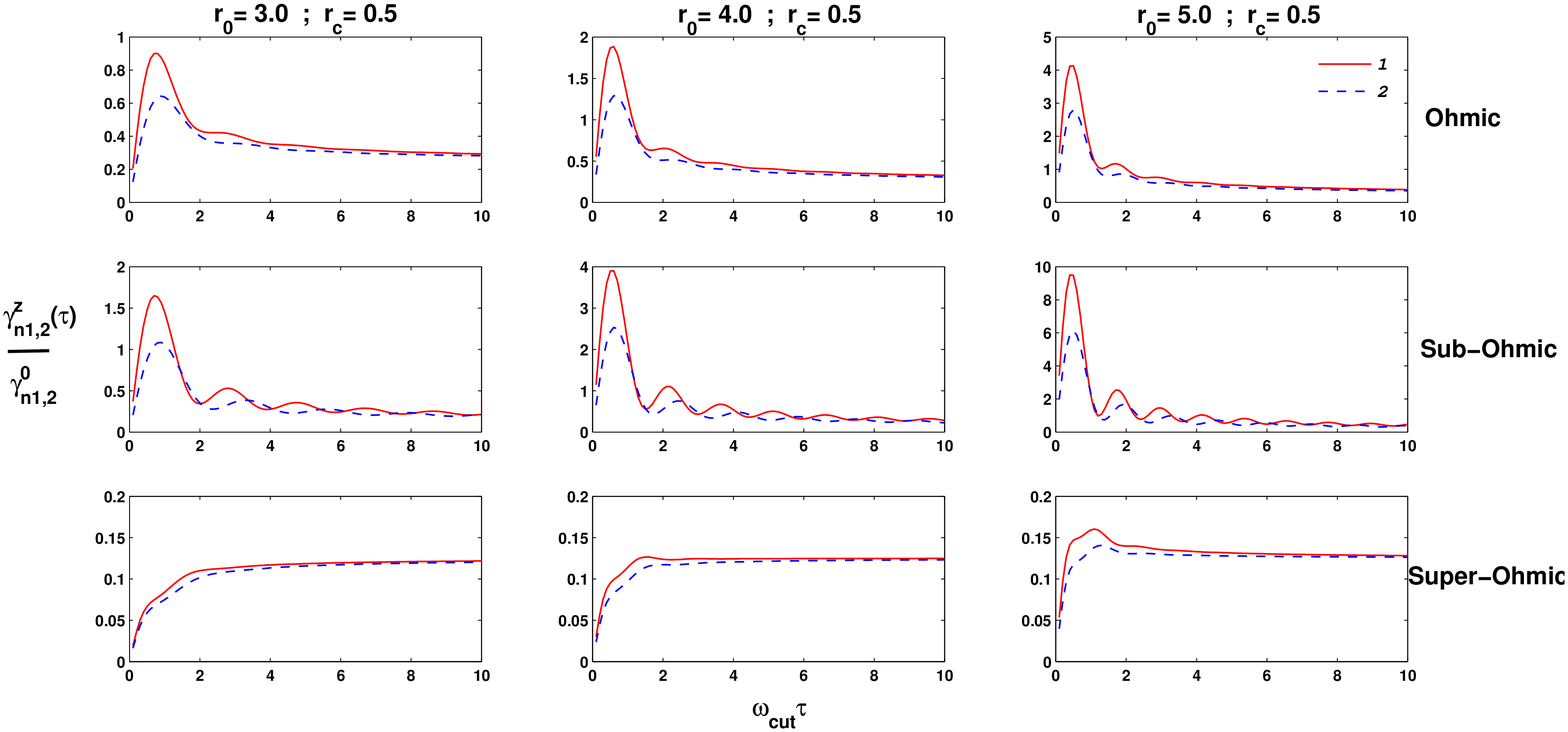}
		\caption{ We plot the ratio between the effective decay rate $\gamma_{n_{1,2}}^z(\tau)$ and the Markovian decay rate $\gamma_{n_{1,2}}^0$  as a function of unitless time $\omega_{1,2}t$  for the Ohmic (uppermost row), sub-Ohmic (middle row) and super-Ohmic (lowermost row) reservoirs at high temperatures with a fixed value of $r_c=0.5$ and by varying confinement length parameter $r_0$. In the regime $r_{1,2}>>1$, we use $\omega_{cut}=1$ and the left, middle and right columns denote $r_0=3.0$, $r_0=4.0$, and $r_0=5.0$ cases, respectively.}
		\end{center}
	\end{figure*}
}

\newcommand{\figfive}
{
	\begin{figure*}[h]
		\begin{center}
		\includegraphics[width=7.0in,height=3.5in]{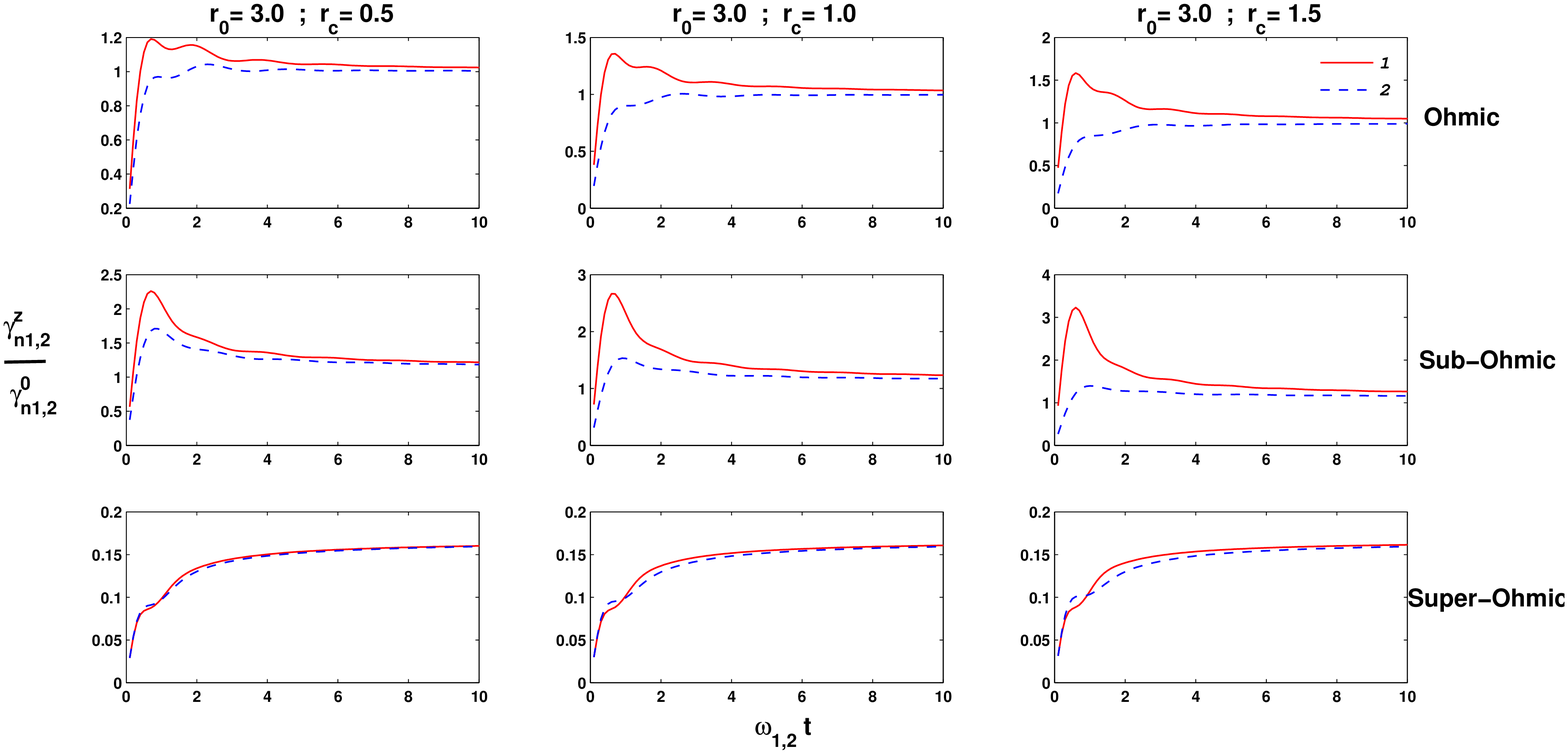}
		\caption{We plot the dynamics of the ratio between the effective decay rate $\gamma_{n_{1,2}}^z(\tau)$ and the Markovian decay rate $\gamma_{n_{1,2}}^0$ as a function of unitless time $\omega_{1,2}t$  for the Ohmic (uppermost row), sub-Ohmic (middle row) and super-Ohmic (lowermost row) reservoirs at zero temperature. We keep fix the value of $r_0=3.0$,and $\omega_{cut}=1$ for the $r_{1,2}>>1$ regime, and vary the magnetic field parameter $r_c$. The three respective columns denote $r_c=0.5$ (left) , $r_c=1.0$ (middle), and $r_c=1.5$ (right) cases, respectively. }
		\label{fig:LTrcrg1}
		\end{center}
	\end{figure*}
}
\newcommand{\figsix}
{
	\begin{figure*}[h]
		\begin{center}
		\includegraphics[width=7.0in,height=3.5in]{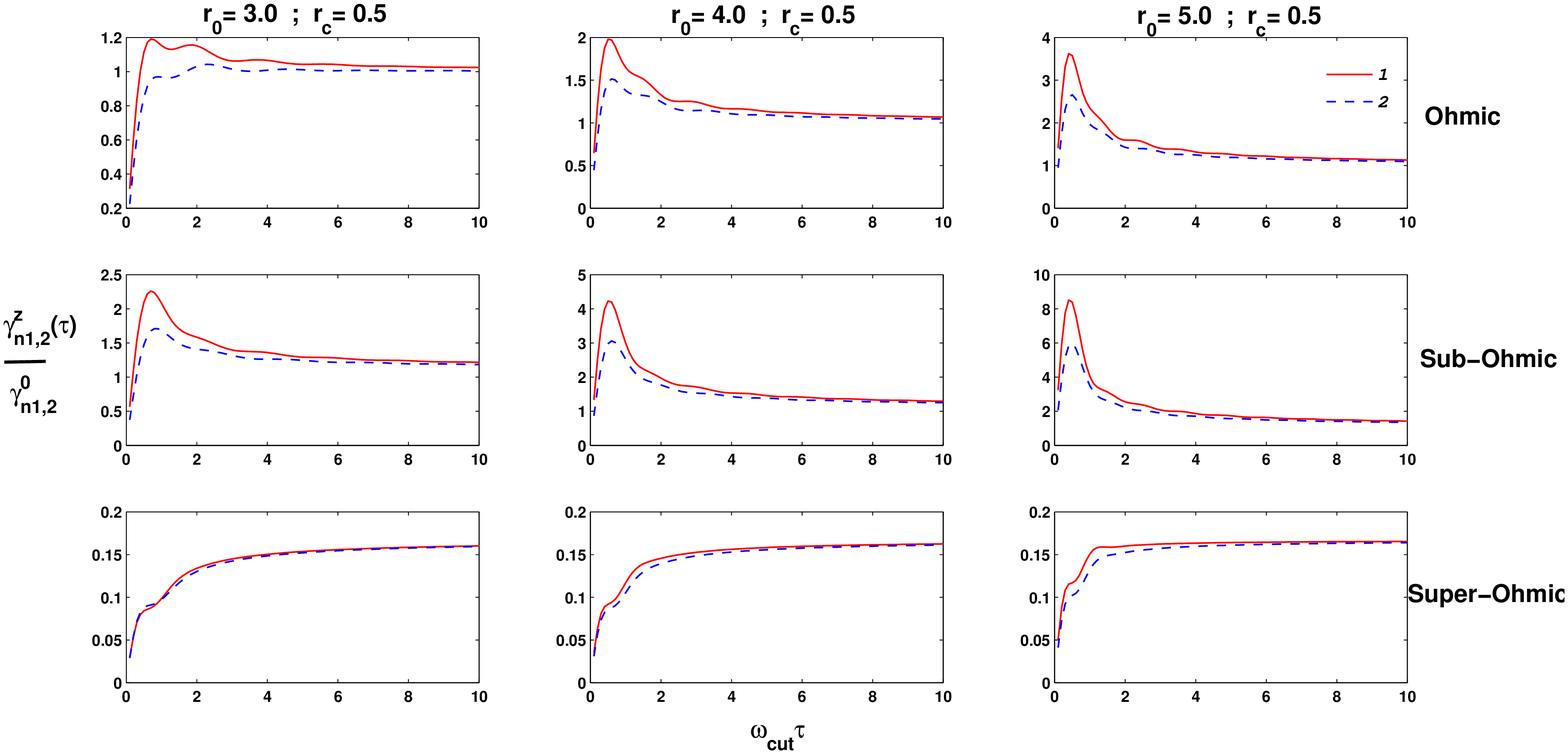}
		\caption{ We plot the ratio between the effective decay rate $\gamma_{n_{1,2}}^z(\tau)$ and the Markovian decay rate $\gamma_{n_{1,2}}^0$  as a function of unitless time $\omega_{1,2}t$  for the Ohmic (uppermost row), sub-Ohmic (middle row) and super-Ohmic (lowermost row) reservoirs at zero temperature with a fixed value of $r_c=0.5$ and $\omega_{cut}=1$. We keep varying confinement length parameter $r_0$ and the left, middle and right columns denote $r_0=3.0$ , $r_0=4.0$, and $r_0=5.0$ cases, respectively. }
		\end{center}
	\end{figure*}
}
\newcommand{\figseven}
{
	\begin{figure*}[h]
		\begin{center}
		\includegraphics[width=7.0in,height=3.5in]{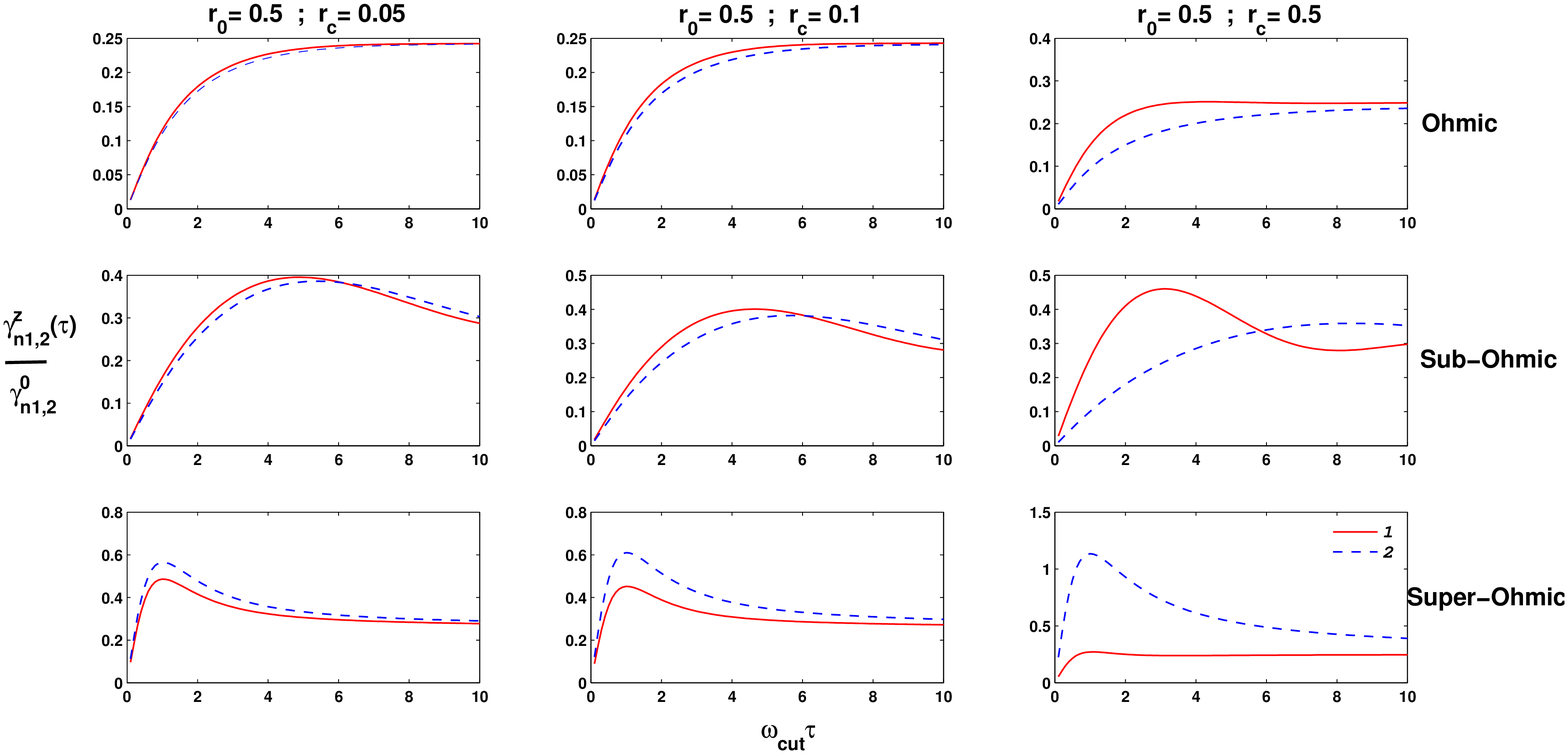}
		\caption{ Dynamics of the ratio between the effective decay rate $\gamma_{n_{1,2}}^z(\tau)$ and the Markovian decay rate $\gamma_{n_{1,2}}^0$ are plotted as a function of unitless time $\omega_{1,2}t$  for the Ohmic (uppermost row), sub-Ohmic (middle row) and super-Ohmic (lowermost row) reservoirs at high temperatures. Here, we keep fix the value of $r_0=0.5$ and change the  magnetic field $r_c$. We use $\omega_{cut}=1$ for the $r_{1,2}<1$ regime, where the left, middle and right columns denote $r_c=0.05$ , $r_c=0.1$, and $r_c=0.5$ cases respectively.}
		\end{center}
	\end{figure*}
}

\newcommand{\figeight}
{
	\begin{figure*}[h]
		\begin{center}
		\includegraphics[width=7.0in,height=3.5in]{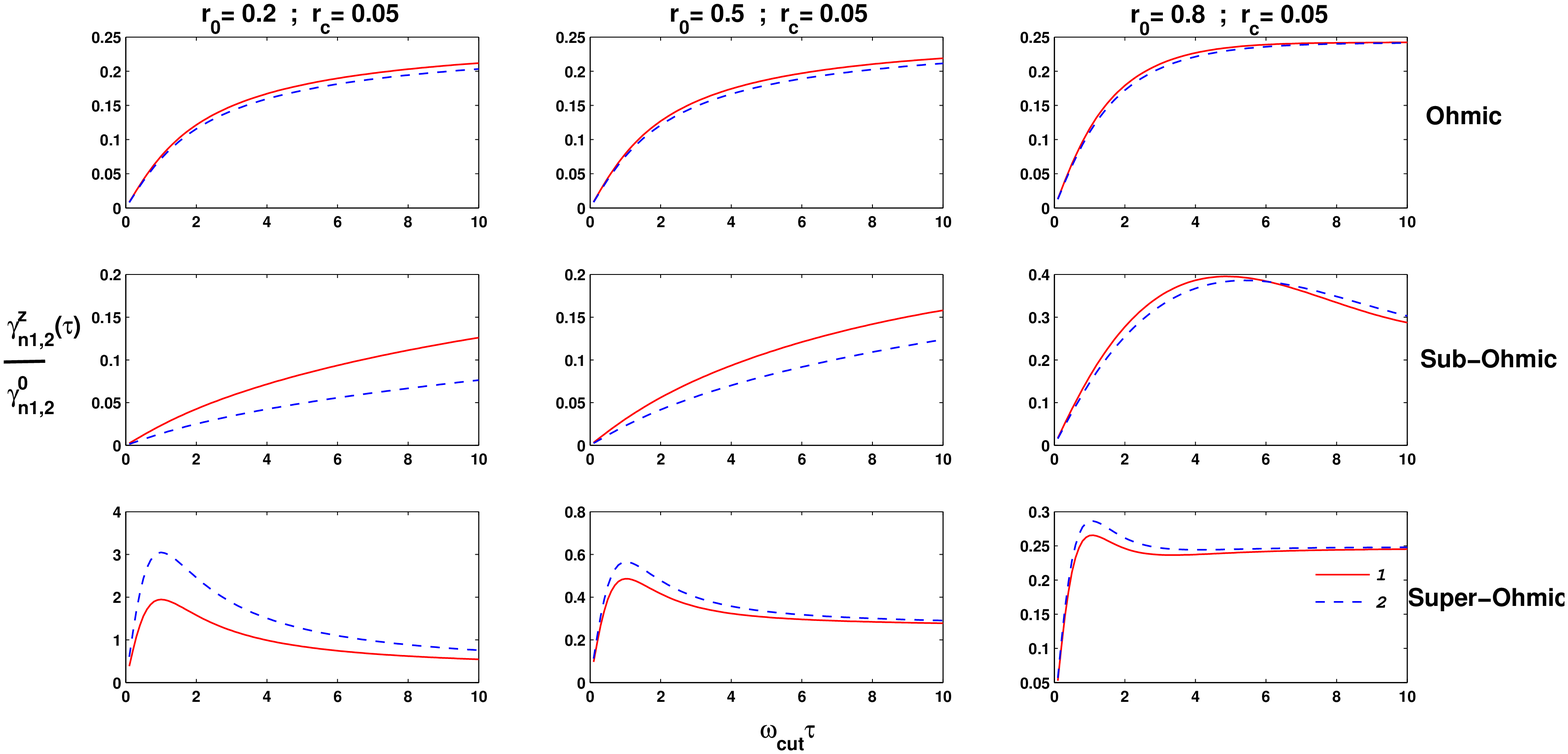}
		\caption{ We plot the ratio between the effective decay rate $\gamma_{n_{1,2}}^z(\tau)$ and the Markovian decay rate $\gamma_{n_{1,2}}^0$  as a function of unitless time $\omega_{1,2}t$  for the Ohmic (uppermost row), sub-Ohmic (middle row) and super-Ohmic (lowermost row) reservoirs at high temperatures with a fixed value of $r_c=0.05$ and $\omega_{cut}=1$. We vary the confinement length parameter $r_0$ and the three respective columns denote $r_0=0.2$ (left), $r_0=0.5$ (middle), and $r_0=0.8$ (right) cases, respectively. }
		\label{fig:HTr0rg1}
		\end{center}
	\end{figure*}
}

\newcommand{\fignine}
{
	\begin{figure*}[h]
		\begin{center}
		\includegraphics[width=7.0in,height=3.5in]{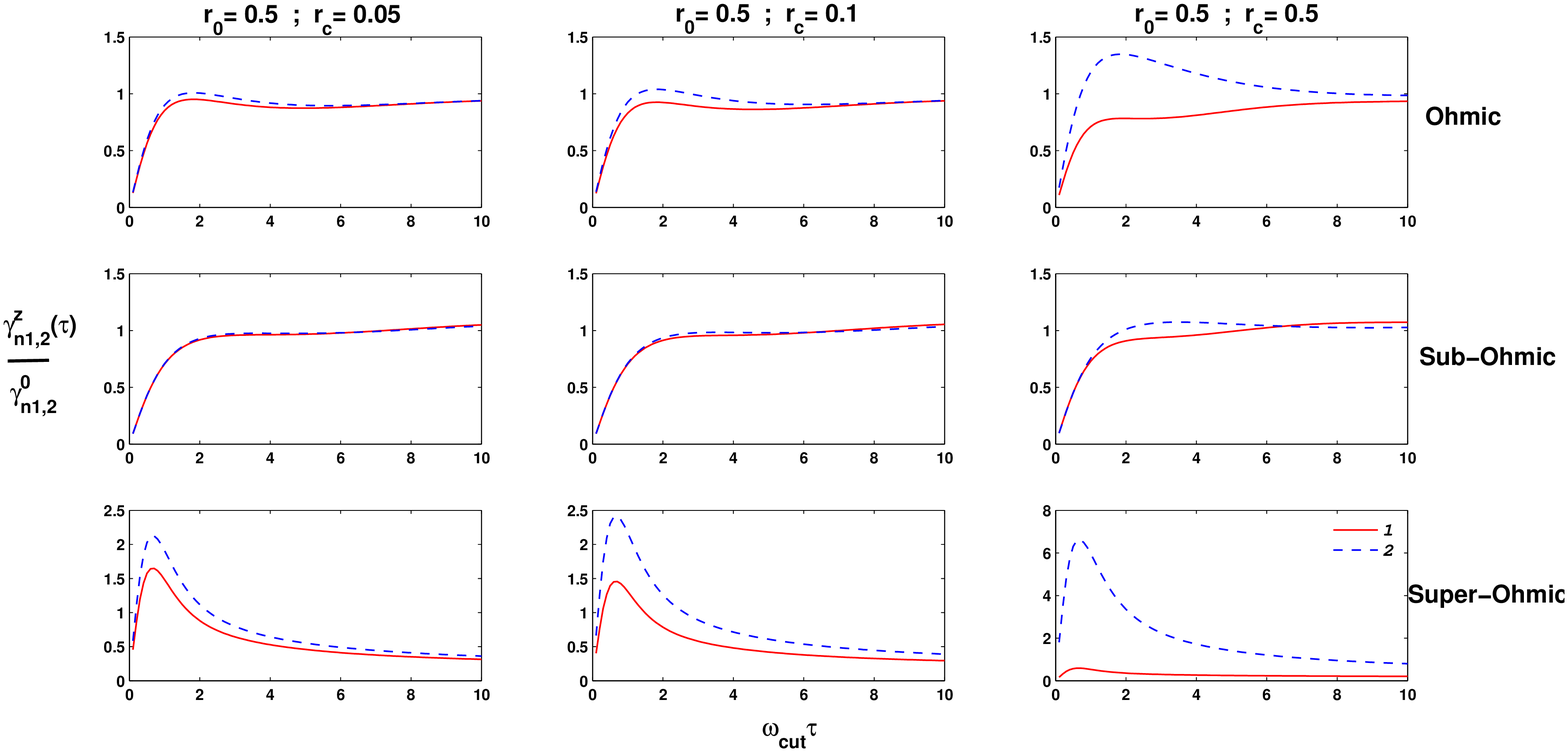}
		\caption{Dynamics of the ratio between the effective decay rate $\gamma_{n_{1,2}}^z(\tau)$ and the Markovian decay rate $\gamma_{n_{1,2}}^0$ are plotted as a function of unitless time $\omega_{1,2}t$  for the Ohmic (uppermost row), sub-Ohmic (middle row) and super-Ohmic (lowermost row) reservoirs at zero temperature. We affix the value of $r_0=0.5$, $\omega_{cut}=1$ for the $r_{1,2}<1$ regime. Here, left, middle, and right columns denote $r_c=0.05$, $r_c=0.1$, and $r_c=0.5$ cases, respectively. }
		\end{center}
	\end{figure*}
}
\newcommand{\figten}
{
	\begin{figure*}[h]
		\begin{center}
		\includegraphics[width=7.0in,height=3.5in]{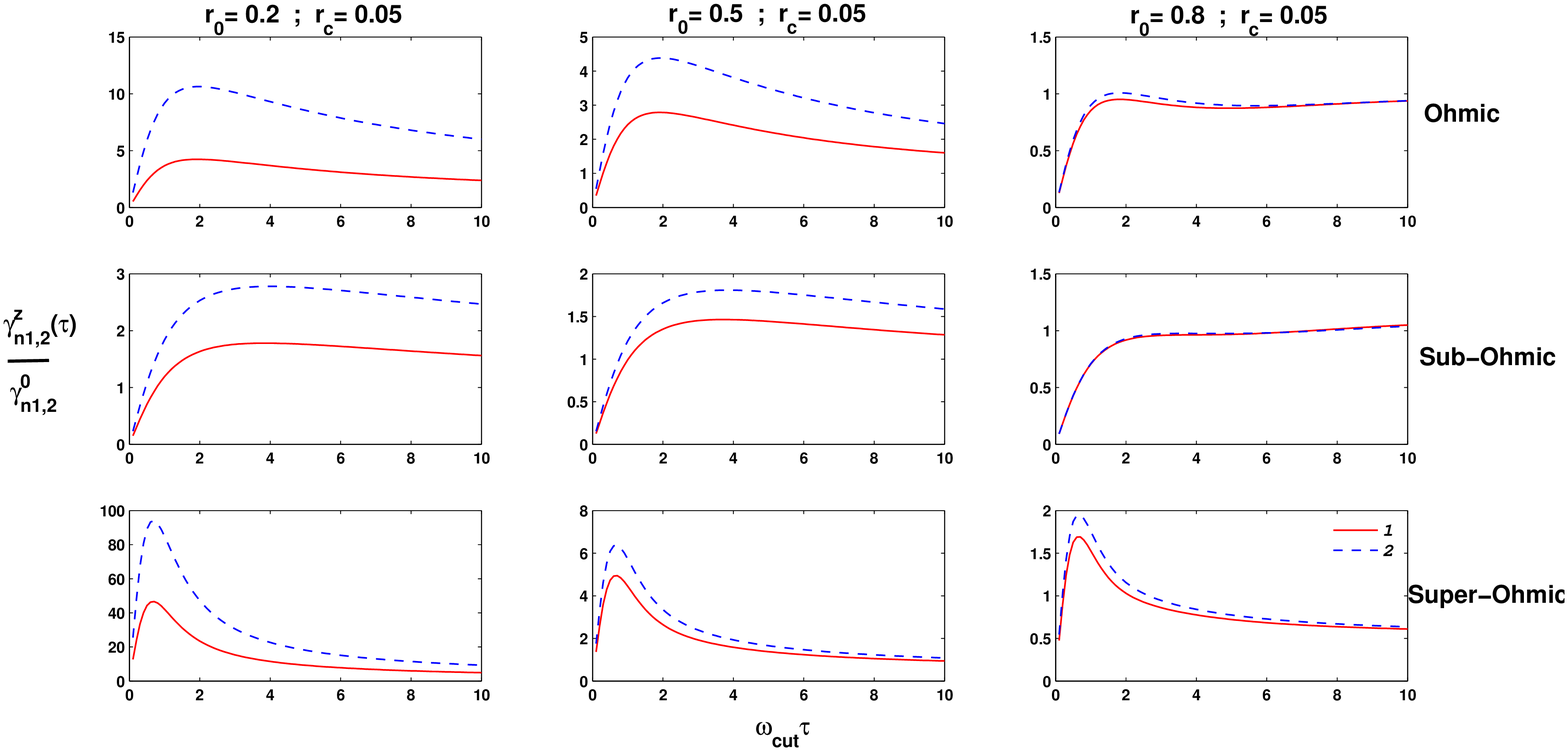}
		\caption{ We plot the ratio between the effective decay rate $\gamma_{n_{1,2}}^z(\tau)$ and the Markovian decay rate $\gamma_{n_{1,2}}^0$  as a function of unitless time $\omega_{1,2}t$  for the Ohmic (uppermost row), sub-Ohmic (middle row) and super-Ohmic (lowermost row) reservoirs at zero temperature with a fixed value of $r_c=0.05$ and $\omega_{cut}=1$ for the $r_{1,2}>1$ regime. The three columns denote $r_0=0.2$ (left) , $r_0=0.5$ (middle), and $r_0=0.8$(right) cases, respectively. }
		\end{center}
	\end{figure*}
}
\newcommand{\figeleven}
{
	\begin{figure*}[t]
		\begin{center}
		\includegraphics[width=6.0in,height=2.0in]{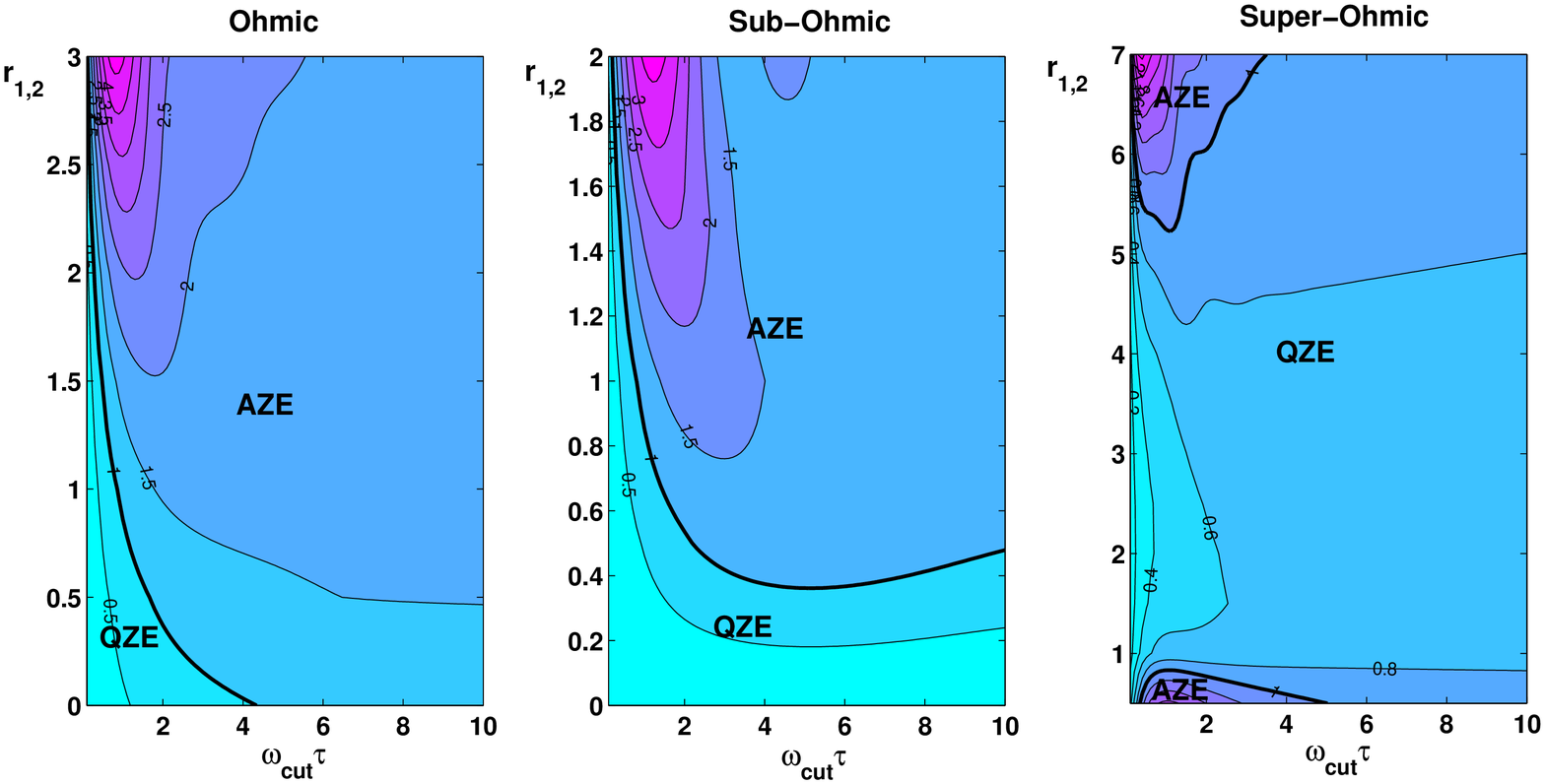}
\caption{ We plot the ratio between the effective decay rate $\gamma_{n_{1,2}}^z(\tau)$ and the Markovian decay rate $\gamma_{n_{1,2}}^0$  as a function of measurement time interval $\omega_{1,2}\tau$ and the parameter $r_{1,2}$ for the Ohmic (left column), sub-Ohmic (middle column) and super-Ohmic (right column) reservoirs at high temperatures. The crossover between QZE and AZE are denoted by bold solid lines.}
		\end{center}
	\end{figure*}
}

\begin{document}
	\title {Control of Decoherence in different environments : A case study for dissipative magneto-oscillator}
	\author {Asam Rajesh$^1$, Malay Bandyopadhyay$^{1,\dagger}$, and A. M. Jayannavar$^{2,3}$}
	\affiliation{1. School of Basic Sciences,Indian Institute of Technology Bhubaneswar,Bhubaneswar 751007,India\\
2. Institute of Physics, Sachivalaya Marg, Sainik School PO, Bhubaneswar, India, 751005.\\
3.Homi Bhabha National Institute, Training School Complex, Anushakti Nagar,
Mumbai-400085, India}
\begin{abstract}
In this paper, we analyze two different techniques based on reservoir engineering method and quantum Zeno effect for controlling decoherence of a dissipative charged oscillator in the presence of an external magnetic field. Our main focus is to investigate the sensitiveness of these decoherence control techniques on the details of different environmental spectrum ($J(\omega)$), and on the crucial role played by different system and reservoir parameters, e.g., external magnetic field ($r_c$), confinement length ($r_0$), temperature ($T$), cut-off frequency of reservoir spectrum ($\omega_{cut}$), and measurement interval ($\tau$). First, we consider the charged quantum oscillator in an initial nonclassical Schr$\ddot{o}$dinger cat state and analyze the non-Markovian dynamics for the magneto-oscillator in contact with Ohmic, sub-Ohmic, and super-Ohmic environments. We show the procedure to control the quantumness of the Schr$\ddot{o}$dinger cat state  by tuning the parameters $r_c$, $r_0$, and $J(\omega)$. On the other hand, we investigate the effect of nonselective energy measurement process on the mortification of quantumness of an initial Fock-Darwin state of the charged magneto-oscillator. We investigate in details the strategy to manipulate the continuous passage from decay suppression to decay acceleration by engineered reservoirs and by tuning the system or reservoir parameters, e.g., $r_c$, $r_0$, $T$ or $\tau$. As a result of that one can control environment induced decoherence (EID).
\end{abstract}
\maketitle
\section{Introduction}
The study of open quantum system has been received enormous attention due to its ubiquitous application in  developing quantum information devices \cite{1,2}, quantum computation, cryptography and reaction yields. These are based on two basic principle of quantum mechanics, i.e.,quantum entanglement and superposition of quantum states. Entanglement and superposition deteriorate when a quantum system interacts with a environment. This coupling transforms the quantum superposition in to classical statistical mixtures which is named as decoherence \cite{3}.  Decoherence is basically involve with the non-unitary evolution of states which has serious consequences like a loss of information or probability leakage of information in environment. The essential ingredient for the efficiency, speed, and security of quantum algorithms and cryptography is the
principle of superposition of states. As mentioned by Unruh \cite{unruh}, the loss of purity of states (known as decoherence) would fall the performance, particularly in the case of long-distance communications and large
scale computations . Thus, the information carried by a quantum bit has to be protected from decoherence.  Naturally, the million dollar question is can quantum systems be controlled and made to outperform the classical systems inspite of decoherence? The answer is affirmative and the great increase in ability to coherent control and manipulation of the state of quantum systems expedite the path to control the transition from quantum to classical world. In the context of coherent control of EID, a large number of investigations are made both in the connection of fundamental issues of quantum theory and in the relation of emerging quantum devices\cite{3,4,5,6}. For instance, one can mention that controlling EID and eventually halting it is a major challenge in quantum computation where several quantum states are required to be specified by a single wave function and information processing are carried out in parallel by unitary operations. \\
\indent
 The fragile nature of quantum superpositions and entangled states due to EID lead us to introduce several methods to protect this delicate but potentially very powerful technique. For instance one can mention the methods based on decoherence free subspaces, dynamical decoupling  and bang bang techniques \cite{viola}. Recently the connection between these techniques and the quantum Zeno effect has been clarified \cite{facchi}. Usually, the study of Zeno and anti-Zeno dynamics is done by assuming that the system is initially prepared in an eigenstate of the free Hamiltonian. Such kind of situations are considered in Refs. \cite{manis1,manis2}. In the present paper our goal is to find  whether the quantum Zeno effect can be used to inhibit quantum decoherence when the system is initially prepared in Schrodinger cat state. In the context of damped harmonic oscillator, Maniscalco {\it et al} explored the possibility of modifying the quantum-classical transition as a consequence of measurements performed on the system through the analysis of Zeno and anti-Zeno phenomena. Further, we introduce an external magnetic field and study its effect in the context of quantum-classical transition. Basically, we study how to modify quantum-classical transition by using reservoir engineering technique for an initial Schrodinger cat state and by using selective energy measurement method for an initial Fock-Darwin state of the charged magneto-oscillator. In the first case, we make a comparative study of three different type of reservoirs (Ohmic, sub-Ohmic and super-Ohmic) and reveal which reservoir succeed in leading slowest decoherence. In the latter case, one can find out that the nonselective measurement may slow down quantum-classical transition (Zeno effect) or may accelerate the transition (anti-Zeno effect). The other aspect which is investigated in the present paper is the connection between the dynamics of the system in the presence of nonselective measurements and the modulation dynamics by controlling the system and reservoir parameters.\\
\indent
\figone
\figtwo
Now, one can summarize the main findings of this paper. We reveal the crucial role played by the environmental spectrum, external magnetic field and the confining length in controlling decoherence properties of a quantum system. We explicitly show that by modifying reservoir spectrum or the external magnetic field or the confining length one can prolong or reduce the life of a Schrodinger cat state. On the other hand, we also demonstrate how the nonselective energy measurements can modify the quantumness of the initial Fock-Darwin state of a damped magneto-oscillator. In the latter case, it is observed that the crossover from Zeno to anti-Zeno is highly sensitive to the tuning parameters like external magnetic field ($r_c$) or the confining length ($r_0$) or the details of reservoir spectrum $J(\omega)$ or the measurement interval $\tau$. For certain type of Bosonic reservoirs, even a very small variation of $r_c$, $r_0$ or $\tau$ can cause a measurement induced acceleration or deceleration of decoherence. \\
\indent
The structure of the paper is as follows. As the structured reservoirs are generally characterized by memory term, one requires a non-Markovian theoretical description to describe coherence properties of a system. In section II, we describe our system and the non-Markovian master equation to investigate the dynamics of the system in contact with different types of reservoir. In section III, we employ reservoir engineering technique to modify the decoherence properties of the above mentioned system and compare the dynamics of an initial Schrodinger cat state for three different types of bosonic environment (Ohmic, sub-Ohmic and super-Ohmic). In section IV, we demonstrate the possibility of controlling quantumness of an initial Fock-Darwin state of the damped magneto-oscillator by using the quantum Zeno or anti-Zeno protocol. Finally, we summarize our results and conclude in section V.
\indent
\section{The system}
 We consider a quantum charged  harmonic oscillator in the presence of an external magnetic field along $z$ direction and linearly coupled to two independent reservoirs which are modelled as a collection of an infinite chain of independent quantum harmonic oscillators. The total Hamiltonian of the system is given by :
 \begin{equation}
 H=H_S+H_B+H_{SB},
 \end{equation}
 where, the system Hamiltonian is
 \begin{equation}
 H_s=\frac{(\vec{p}-\frac{e\vec{A}}{c})^2}{2m}+\frac{1}{2}m\omega_0^2(x^2+y^2),
 \end{equation}
 with $\vec{B}=\vec{\nabla}\times\vec{A}$; $\vec{A}=\frac{B_0}{2}(-y,x,0)$. One can recast Eq. (2) as follows :
  \begin{equation}
 H_s=(a_1^{\dagger}a_1+\frac{1}{2})\hbar\omega_1 + (a_2^{\dagger}a_2+\frac{1}{2})\hbar\omega_2,
 \end{equation}
 with $\omega_{1,2}=\omega^{\prime}\pm \frac{\omega_c}{2}$, the cyclotron frequency $\omega_c=\frac{eB_0}{mc}$ and $\omega^{\prime}=\sqrt{\omega_0^2+\omega_c^2/4}$. Here, $a_{1,2}^{\dagger}= \Big\lbrack \frac{(x\pm iy )}{2l}-\frac{l}{2}(\partial_x\pm i\partial_y)\Big\rbrack$, and  $a_{1,2}=\Big\lbrack \frac{(x\mp iy)}{2l}-\frac{l}{2}(\partial_x\mp i\partial_y)\Big\rbrack$, are the Fock-darwin creation and annihilaton operators for the two independent oscillators (denoted by 1 and 2 ) of the charged magneto-oscillator, and $l=\sqrt{\frac{\hbar}{m\omega^{\prime}}}$ is the Fock-Darwin radius.\\
 \indent
 The environment is  modeled as an infinite chain of harmonic oscillators. In the present context, we consider two independent heat baths in the x and y direction respectively :
 \begin{equation}
 H_B= \sum_{n=1,k=1,2}^{n=\infty}\hbar\omega_{n,k}(b_{n,k}^{\dagger}b_{n,k}+\frac{1}{2}),
 \end{equation}
 where $b_{n,1}$ and $b_{n,1}^{\dagger}$ are the annihilation and creation operators of heat bath oscillators in the x direction, respectively and $\omega_{n,1}$ is the frequency of the $n^{th}$ oscillator in the x-direction. The same things are applicable to the y-direction heat bath with the replacement of 1 by index 2. The system and the reservoirs are coupled linearly via position type operators $x=\frac{l}{2}\lbrack (a_1+a_1^{\dagger})+(a_2+a_2^{\dagger})\rbrack$, and $y=-i\frac{l}{2}\lbrack (a_1^{\dagger}-a_1)-(a_2^{\dagger}-a_2)\rbrack$ and $x_{n,k}\propto (b_{n,k}+b_{n,k}^{\dagger})$ for the system and reservoir oscillators respectively. This is just to remind you that the interaction may result in position-momentum or momentum-position type coupling and they are equivalent to position-position coupling which is explicitly discussed in Ref. \cite{ford}. Now, the interaction Hamiltonian is given by :
 \begin{eqnarray}
 H_{SB}=\frac{l}{2\sqrt{2}}\Big\lbrack(a_1^{\dagger}+a_1)+(a_2^{\dagger}+a_2)\Big\rbrack\sum_n c_{n,1}(b_{n,1}+b_{n,1}^{\dagger})\nonumber \\
 \frac{l}{2i\sqrt{2}}\Big\lbrack(a_1^{\dagger}-a_1)-(a_2^{\dagger}-a_2)\Big\rbrack\sum_n c_{n,2}(b_{n,2}+b_{n,2}^{\dagger}),
 \end{eqnarray}
 where, $c_{n,k}$ describes the interaction strength between the system with each mode of the reservoir in the kth direction (remember k=1 indicates the reservoir in the x direction and k=2 is for the y direction reservoir). From the above analysis one can recast our model in terms of two independent harmonic oscillators coupled to two independent oscillators. In the next subsection, we describe the non-Markovian master equation derived from the total Hamiltonian as described in Eq. (1).
 \figthree
 \figfour
  \subsection{Non-Markovian Master Equation}
  In this section we describe the general derivation of the quantum master equation for the reduced density matrix starting from the microscopic  Hamiltonian as described in the previous section. Here, we follow the derivation methods as described in \cite{breuer}. We assume that the system and environment are weakly coupled. The derivation basically starts from the interaction picture von Neumann equation \begin{equation}
  \frac{d\tilde{\rho}(t)}{dt}=-i\lbrack \tilde{H}_{SB},\tilde{\rho}(t)\rbrack,
  \end{equation}
  with $\tilde{\rho}(t)=e^{it(H_S+H_B)}\rho(t)e^{-it(H_S+H_B)}$. Integrating the above expression, we can obtain a Dyson series and one can safely neglect the terms that are higher than the second order in the coupling strength. Now, tracing over the bath degrees of freedom one can obtain an equation for the reduced system. To proceed further, one needs to assume few assumptions. First, one can assume that the system and environment are uncorrelated at the initial time $t=0$. Secondly, we consider the environment as stationary. The master equation can be simplified by considering the thermal reservoir at temperature $T$ described by a density operator of the form :
  \begin{equation}
  \rho_{B_k}=\frac{1}{Z_{B_k}}\exp\Big(- \sum_{n=1}^{\infty}\frac{\omega_{n_k}b_{n_k}^{\dagger}b_{n_k}}{k_BT}\Big),
  \end{equation}
  where $k_B$ is the Boltzmann constant and $Z_{B_k}$ is the partition function of the independent bath in the kth direction (k=1 denotes x direction and k=2 is for y direction). As a result of this one can obtain $Tr_{B_k}\lbrace\lbrack\tilde{H}_{SB_k}(t),\tilde{\rho}_S(t)\otimes \rho_{B_k}(t)\rbrack\rbrace=0$. Therefore one can obtain the following simplified master equation :
  \begin{equation}
  \frac{d\tilde{\rho}_S(t)}{dt}=-\alpha^2\int_{0}^{t}dt^{\prime} Tr_{B}\lbrace\lbrack \tilde{H}_{SB}(t), \lbrack\tilde{H}_{SB}(t^{\prime}),\tilde{\rho}_S(t)\otimes \rho_{B}(t)\rbrack\rbrack\rbrace\rbrack.
  \end{equation}
  Now, as described in Ref.\cite{breuer}, one can introduce the eigenoperators of the system Hamiltonian and can express the interaction Hamiltonian $H_{SB}$ in terms of these eigenoperators. As a result of that the interaction Hamiltonian in the interaction picture can be written as follows :
  \begin{eqnarray}
  H_{SB}(t)=\frac{l}{2\sqrt{2}}\Big\lbrack \tilde{A}_1(t)+ \tilde{A}_2(t)\Big\rbrack\otimes \tilde{E}(t)+
 \frac{l}{2i\sqrt{2}}\Big\lbrack \tilde{B}_1(t)-\tilde{B}_2(t)\Big\rbrack\otimes \tilde{F}(t),
  \end{eqnarray}
  where, $\tilde{A}_1(t)=e^{-i\omega_1t}a_1+e^{i\omega_1t}a_1^{\dagger}$ , $\tilde{A}_2(t)=e^{-i\omega_2t}a_2+e^{i\omega_2t}a_2^{\dagger}$,$\tilde{B}_1(t)=e^{-i\omega_1t}a_1- e^{i\omega_1t}a_1^{\dagger}$ , $\tilde{B}_2(t)=e^{-i\omega_2t}a_2-e^{i\omega_2t}a_2^{\dagger}$, $\tilde{E}(t)=\sum_n c_{n,1}(b_{n,1}(t)+b_{n,1}^{\dagger}(t))$ and $\tilde{F}(t)=\sum_n c_{n,2}(b_{n,2}(t)+b_{n,2}^{\dagger}(t))$. Now, using the definition of a correlation function for the field operators $\tilde{E}(t)$ as $<\tilde{E}(t)\tilde{E}(t^{\prime})>=Tr_B\lbrace \tilde{E}(t)\tilde{E}(t^{\prime})\rho_B  \rbrace$. Now,
   one can obtain a non-Markovian master equation describing the system dynamics \cite{manis3} :
  \begin{widetext}
  \begin{eqnarray}
    \frac{d\rho_s(t)}{dt}&=& \frac{\Delta_1 + \Gamma_1}{2}\lbrack L(a_1)+L(a_2,a_1)\rbrack+\frac{\Delta_1 - \Gamma_1}{2}\lbrack L(a_1^{\dagger})+L(a_2^{\dagger},a_1^{\dagger})\rbrack \nonumber \\
    &+&\frac{\Delta_2 + \Gamma_2}{2}\lbrack L(a_2)+L(a_1,a_2)\rbrack+\frac{\Delta_2 - \Gamma_2}{2}\lbrack L(a_2^{\dagger})+L(a_1^{\dagger},a_2^{\dagger})\rbrack \nonumber \\
     &+&\frac{\Delta_1+\Gamma_1}{2}\lbrack e^{(2i\omega_2t)} D(a_1)+D(a_1,a_2)\rbrack+\frac{\Delta_1-\Gamma_1}{2}\lbrack e^{(2i\omega_1t)} D(a_1^{\dagger})+D(a_1^{\dagger},a_2^{\dagger})\rbrack \nonumber \\
     &+&\frac{\Delta_2+\Gamma_2}{2}\lbrack e^{(2i\omega_1t)} D(a_2)+D(a_2,a_1)\rbrack+\frac{\Delta_2-\Gamma_2}{2}\lbrack e^{(2i\omega_2t)} D(a_2^{\dagger})+D(a_2^{\dagger},a_1^{\dagger})\rbrack \nonumber \\
     \end{eqnarray}
  where, the super-operators $L(O)$, $D(O)$, $L(O_1,O_2)$ and $D(O_1,O_2)$ are given by :
  \begin{eqnarray}
  L(O)=2O^{\dagger}\rho O-OO^{\dagger}\rho-\rho OO^{\dagger} \nonumber \\
  D(O)=2O\rho O-O^2\rho-\rho O^2 \nonumber \\
  L(O_1,O_2)=2O_1^{\dagger}\rho O_2-O_1O_2^{\dagger}\rho-\rho O_1O_2^{\dagger} \nonumber \\
  D(O_1,O_2)=2O_1\rho O_2-O_1O_2\rho-\rho O_1O_2 .
  \end{eqnarray}
\end{widetext}
  Since, the decoherence occurs mostly at time scales much shorter than $t_{th}$, one can use the master equation (8) for the entire paper. Later on many results will be discussed for the high temperature regime, and in the deep classical regime where thermal energy $k_BT$ is hundred times greater than the oscillator energies $\hbar\omega_{1,2}$. One can easily express the thermalization time at high temperatures in terms of $\omega_{1,2}$ as follows \cite{manis1}:
  \begin{equation}
  t_{th}^{1,2}=\frac{\omega_{1,2}}{\Gamma}=\frac{1}{\pi\alpha^2}r_{1,2}^{s-1}\exp(1/r_{1,2})
  \end{equation}
Thus, one can say that the thermalization time,$t_{th}^{1,2}$, decreases monotonically for increasing values of $r_{1,2}$ (this resonance parameter defined later in Eq. ) for both the sub-Ohmic and the Ohmic reservoirs. This implies that $t_{th}^{1,2}$ decreases with the decreasing values of the cutoff frequency with respect to the frequency of the system oscillators i.e. $\omega_{1,2}$. On the other hand, there exist a value of $r_{1,2}$ which minimizes the thermalization time for the super-Ohmic reservoir. In general, the thermalization time increases rapidly as $r_{1,2}\rightarrow 0$ for all the three reservoir types considered in this paper. As a matter of fact, the thermalization process is notably slowed down. Thus, one can control the thermalization dynamics,
by appropriately changing the cutoff frequency of a high temperature engineered reservoir and also varying the oscillators frequencies $\omega_{1,2}$.
  \indent
   Assuming an initially factorized state $(\rho=\rho_s\otimes\rho_B)$ ($\rho_B$ is the density operator of the Bath) with the weak coupling limit and secular approximation, performing a coarse grain over time scales of the order of $\frac{1}{\omega_{1,2}}$, one can  obtain the following approximated master equation \cite{5}:
   \begin{widetext}
  \begin{eqnarray}
  \frac{d\rho_s(t)}{dt}&= &\frac{\Delta_1 + \Gamma_1}{2}\lbrack L(a_1)+L(a_2,a_1)\rbrack+\frac{\Delta_1 - \Gamma_1}{2}\lbrack L(a_1^{\dagger})+L(a_2^{\dagger},a_1^{\dagger})\rbrack \nonumber \\
    &+&\frac{\Delta_2 + \Gamma_2}{2}\lbrack L(a_2)+L(a_1,a_2)\rbrack+\frac{\Delta_2 - \Gamma_2}{2}\lbrack L(a_2^{\dagger})+L(a_1^{\dagger},a_2^{\dagger})\rbrack
  \end{eqnarray}
  \end{widetext}
  This is to inform you that $\Delta_{1,2}=\Delta_{1,2}^x + \Delta_{1,2}^y$
  One can define the coefficients as follows :
  \begin{eqnarray}
 &&\Sigma_{1,2}=\frac{\Delta_{1,2} + \Gamma_{1,2}}{2}, \\
  &&\Lambda_{1,2}=\frac{\Delta_{1,2}-\Gamma_{1,2}}{2}, \\
 &&\Delta_{1,2}(t)=4\int_0^t dt^{\prime}\int d\omega J(\omega)\Big[N(\omega)+\frac{1}{2}\Big]\cos(\omega t^{\prime})\nonumber \\
  &&\times\cos(\omega_{1,2} t^{\prime}), \\
  &&\Gamma_{1,2}(t)=4\int_0^t dt^{\prime}\int d\omega \frac{J(\omega)}{2}\sin(\omega t^{\prime})\sin(\omega_{1,2} t^{\prime}),\nonumber \\
  &&N(\omega,T)=1/\lbrack e^{\hbar\omega/k_BT}-1\rbrack \nonumber \\
  &&J(\omega)=\alpha^2 \sum_n \frac{k_n^2}{m_n\omega_n}\delta(\omega-\omega_n).
  \end{eqnarray}
  Where $\Delta_{1,2}(t)$ and $\Gamma_{1,2}(t)$ are the diffusion and dissipation coefficients for the two modes defined by $1$ and $2$. We follow this nomenclature for the rest of the paper. This is just to inform you that we have used this combined form for other variables. We separate out them by introducing a comma notation between them. Wherever, we use this notation {1,2}, it actually imply to specify the two different modes $1$ and  mode $2$ in the combined form. They are basically denoting two different circularly polarized modes (clockwise and anti-clockwise). This is to mention here that $\Delta_{1,2}=\Delta_{1,2}^x+\Delta_{1,2}^y$ and $\Gamma_{1,2}=\Gamma_{1,2}^x+\Gamma_{1,2}^y$. Since, we are using two independent baths in the x and y direction, the diffusion and dissipation coefficients are additive. $N(\omega,T)$ is the average number of reservoir thermal excitations, with $k_B$ as the Boltzmann constant and $T$ as the reservoir temperature, and $J(\omega)$ is the spectral density of the environment.  Here, we have to remember that in deriving the above mentioned master equation no Markovian approximation is applied. The memory effects are incorporated through the time dependent coefficients $\Delta_{1,2}(t)$ and $\Gamma_{1,2}(t)$. The justification of using the secular approximated master Eq. (11) is further analyzed in Sec. IV.
  \figfive
  \figsix
 \subsection{Modeling the reservoir}
  Now, we introduce a class of spectral densities to derive decoherence dynamics for different types of reservoirs. The spectral densities which we have introduced for the study of decoherence dynamics of magneto-oscillator is given by \cite{5},
  \begin{equation}
  	J(\omega)=\alpha^2\omega_{cut}^{1-s}\omega^se^{-\omega/\omega_{cut}},
  \end{equation}
  where we have introduced exponential cut-off to eliminate divergence problem in the $\omega\rightarrow\infty$ limit, where $\alpha$ is a dimensionless coupling constant and s is a real parameter which can acquire the values $<1$, $1$,  and  $>1$, corresponding to the so-called  sub-Ohmic, Ohmic and super-Ohmic reservoirs, respectively. We consider the values of s as 1/2, 1, and 3 for the sub-Ohmic, Ohmic and super-Ohmic reservoirs, respectively. These three cases describe different physical contexts.\\
  \indent
  Now we introduce the spectral distribution function which can be written as,\\
  \begin{equation}
  I(\omega)=J(\omega)\big[N(\omega)+\frac{1}{2}\big]
  \end{equation}
  This spectral distribution function contains all information about the reservoirs in the weak-coupling limit. At high temperature one can assume $N(\omega)\propto\dfrac{k_B T}{\omega}$ and at zero temperature as $N(\omega)=0$. Now, we introduce a useful resonance parameter,\\
  \begin{equation}
  r_{1,2}=\frac{\omega_{1,2}}{\omega_{cut}}= \sqrt{r_0^2+r_c^2/4}\pm r_c/2,
  \end{equation}
  where $r_0=\frac{\omega_0}{\omega_{cut}}$ and $r_c=\omega_c/\omega_{cut}$. Thus, $r_{1,2}=\sqrt{r_0^2+r_c^2/4}\pm r_c/2$. The environment induced dynamics of the system can be controlled by changing this resonance parameter. One can shift the system oscillator frequency with respect to the reservoir spectrum by changing the resonance parameters $r_{1,2}$. Thus, it allows to control the effective coupling between the system and reservoir for the two different channels mentioned above. When we have $r_{1,2}>> 1$,  the system oscillator oscillates off-resonant with respect to the peak of the reservoir spectrum for the three types of reservoirs considered in this work and is known as off-resonance case.  On the other hand , for $r_{1,2}<<1$, the oscillator frequency is in resonance with respect to the peak of the reservoir spectrum and is known as resonance case. Since, the secular approximation holds good in the off-resonant case, one can use Eq. (11) to study the non-Markovian decoherence dynamics of the oscillator. On the other hand, the secular approximation fails for $r_{1,2}<<1$ case, and we must use the master equation (8) for this purpose.\\
  \indent
  In the following section, we introduce the fringe visibility function to characterize the decoherence properties for the system prepared initially in a Schr$\ddot{o}$dinger cat state.  We will use the above mentioned two non-Markovian master equations (Eq. (8) and (11)), to examine the decoherence properties of a charged magneto-oscillator in contact with three different types of engineered reservoirs.
  \section{Decoherence Control : Reservoir Engineering Technique}
  We consider that the system is initially prepared in a Schr$\ddot{o}$dinger cat state of the following form :
  \begin{equation}
    |\Psi> = \frac{1}{\sqrt{\mathcal{N}}}\Big( |\alpha_1,\alpha_2>+|-\alpha_1,-\alpha_2>\Big),
  \end{equation}
  where, $|\alpha_1,\alpha_2>$ is a coherent state and $\mathcal{N}^{-1}=2\lbrack 1+\exp\Big\lbrace-4(|\alpha_1|^2+|\alpha_2|^2)\Big\rbrace\rbrack$. Here, $\alpha_1=\alpha_1^x+\alpha_1^y$ and $\alpha_2=\alpha_2^x+\alpha_2^y$. This is to be mentioned here that both the oscillators are placed in the same Schr$\ddot{o}$dinger cat state mentioned in Eq. (21). Further, we assume $\alpha_1$ and $\alpha_2$  to be real for simplicity. Now, we want to study short time decoherence dynamics of the above mentioned cat state. This type of states are also known as even coherent state for their properties that the  even components of the number probability distribution are nonvanishing. The signature of nonclassicality of such type of state is the oscillations of number state probability distribution. In the literature (see Ref. \cite{kim} and references therein), an extensive study has been done on the nonclassical properties of even coherent state, such as negativity of the corresponding Wigner function, oscillation of the number probability distribution etc. Such type of even coherent state and its quantum-classical transition has been  experimentally verified in the context of trapped ion\cite{myatt}\\
  \indent
  In order to study the transition from the initial even coherent state to its corresponding statistical mixture due to its interaction with the environment, it is useful to investigate the dynamics of the corresponding Wigner function. Since, the quantum-classical transition is identified by the fast disappearance of the interference fringes of Wigner function, we need to investigate (to study non-Markovian dynamics) the time evolution of the Wigner function for times $t\leq 1/{\omega_{1,2}}$. Now, we investigate analytic solutions for the short time non-Markovian regime in the two limits : off-resonant case ($r_{1,2}>> 1$) and resonant case ($r_{1,2}<< 1$).
  \subsection{The Off-resonant case $r_{1,2}>>1$}
  In the off-resonant regime ($r_{1,2}>>1$), the secularly approximated non-Markovian master equation (8) can be able to describe the non-Markovian short time decoherence dynamics. In Ref. \cite{7}, the solution of Eq. (8) in terms of quantum characteristic function (QCF) $\chi(\xi)$ is derived. Now, One can judiciously write down the Wigner function as the Fourier transform of the QCF for two independent oscillators in common environment :
  \begin{eqnarray}
   W(\beta_1,\beta_2)&=&\frac{1}{\pi^2}\int_{-\infty}^{\infty}\int_{-\infty}^{\infty}d^2\xi_1d^2\xi_2\chi(\xi_1)\chi(\xi_2)\exp\Big(\beta_1\xi_1^* \nonumber \\
   && - \beta_1^*\xi_1\Big)\exp\Big(\beta_2\xi_2^* - \beta_2^*\xi_2\Big),
  \end{eqnarray}
  where, $*$ denotes complex conjugate form. The Wigner function $W(\beta_1,\beta_2)$ for the above mentioned even coherent state is comprised of two Gaussian peaks centred about $\beta_1=(\pm \alpha_1,\pm\alpha_1);\beta_2=(\pm \alpha_2,\pm\alpha_2) $ respectively with an interference term in between the peaks. The appearance of the interference term indicates the quantumness of the state and its disappearance is considered as a signature of quantum to classical transition. This is to be mentioned that the two-mode Wigner distribution function $W(\beta_1,\beta_2)$ for two oscillators can be thought of as a quasiprobability distribution in the four dimentional space. Now, one can introduce the fringe visibility function (FVF) which is an useful quantity to monitor the transition from the initial even coherent state to its corresponding classical statistical mixture due to the interaction with the environment and  FVF is given by \cite{manis1,manis2} :
  \begin{widetext}
\begin{eqnarray}
F(\alpha_1,\alpha_2,t) = \exp(-A_{int})
   = \frac{1}{2}\frac{W_I(\beta_1,\beta_2,t)|_{peak}}{\Big\lbrack W_{(+\alpha_1,+\alpha_2)}(\beta_1,\beta_2,t)|_{peak}W_{(-\alpha_1,-\alpha_2)}(\beta_1,\beta_2,t)|_{peak}\Big\rbrack^{1/2}},
  \end{eqnarray}
  \end{widetext}
where, $W_I(\beta_1,\beta_2,t)|_{peak}$ denotes the value of the Wigner function at $\beta_1=(0,0);\beta_2=(0,0) $. On the other hand, $W_{(\pm\alpha)}(\beta_1,\beta_2,t)|_{peak}$ define the values of the Wigner function at $\beta_1 = (\pm\alpha_1,\pm \alpha_1);\beta_2 = (\pm\alpha_2,\pm\alpha_2)$, respectively.  For our even coherent state, one can obtain the final expression of FVF for the off-resonant case ($r_{1,2}>>1$) :
\begin{eqnarray}
F(\alpha_1,\alpha_2,t)&=&\exp\Big\lbrack -4\lbrace\alpha_1^2+\alpha_2^2\rbrace \nonumber\\ &&\Big(1-\frac{e^{-(\gamma_{+}(t)+\gamma_{+,-}(t))}}{2(N_{+}(t)+N_{+,-}(t))+1}\Big)\Big\rbrack,
\end{eqnarray}
where,
\begin{eqnarray}
N_{1,2}(t)&=&\int_{0}^{t}dt^{\prime}\Delta_{1,2}(t^{\prime}) \\
\gamma_{1,2}(t)&=&2\int_{0}^{t}dt^{\prime}\Gamma_{1,2}(t^{\prime})
\end{eqnarray}
\subsection{The resonant case : $r_{1,2}<<1$}
Now, we are interested to study the dynamics in the opposite regime, i.e., the resonant case ($r_{1,2}<<1$). In this regime, one needs to solve the full non-Markovian master equation (11). To find the Wigner function, one can proceed by following the same method as that of the Markovian case. First of all, One needs to find the Fokker-Planck equation corresponding to the master equation (11). We know that the initial state is a linear combination of Gaussian terms and the structure of our master equation ensures that each Gaussian term evolves independently. Hence, the final evolved state should also be a linear combination of Gaussian terms. In the resonant regime, the FVF becomes \cite{manis1,manis2} :
\begin{eqnarray}
F(\alpha_1,\alpha_2,t)&=&\exp\Big\lbrack -4\lbrace\alpha_1^2+\alpha_2^2\rbrace\nonumber \\
&&\Big(1-\frac{e^{-(\gamma_{1}(t)+\gamma_{2}(t))}}{4(N_{1}(t)+N_{2}(t))+1}\Big)\Big\rbrack,
\end{eqnarray}
If one compares this equation with that of the FVF for the off-resonant regime (Eq. (24)), one can observe that the only difference is the appearance of an extra two factor in front of the mean energy of the oscillator $N_{1}(t)+N_{2}(t)$ (also known as heating function) in the Eq.(27) compare to that of  Eq. (24). This implies that the heating process in the resonant regime can be thought of due to an effective thermal reservoir with double temperature compare to that of the real temperature. This difference arises due to the appearance of counter rotating terms in Eq. (8). These additional terms provide two extra paths for each channel ($1$ and $2$) to exchange its energy with the environment. In the following subsection we analyze the time evolution of FVF for three different type of reservoirs namely, Ohmic, sub-Ohmic and super-Ohmic.
\figseven
\figeight
\subsection{FVF dynamics:Similarities and differences}
Here, we are interested in investigating the FVF dynamics for three different type of reservoirs : Ohmic, sub-Ohmic and super-Ohmic. The time evolution of FVF are plotted in Figs. (1) and (2) for the off-resonant and resonant cases, respectively. Unlike the typical exponential Markovian time evolutian, one can observe initial non-Markovian quadratic behaviour for all the cases. We also make a comparative study of FVF dynamics for three different type of reservoirs. We also plot the Markovian versus non-Markovian fringe visibilities in the insets of Figs. (1) and (2) for the Ohmic reservoir. \\
\indent
First of all, we analyze the Markovian versus non-Markovian dynamics for FVF for the initial short times. It is observed that the Markovian FVF decays slower than that of the non-Markovian case in the off-resonant regime. One can say that the initial startle in $\Delta_{1,2}(t)$ for the non-Markovian case may cause the faster decoherence in the off-resonant regime compare to the Markovian case. On the other hand, the decay of the FVF is faster for the Markovian case than that of the non-Markovian one in the resonant regime. Since, $\Delta_{1,2}(t)<\Delta_M$ in the resonant regime, one can always found a faster decoherence for the Markovian system. It is just to inform you that the other type of reservoirs (like sub-Ohmic and super-Ohmic reservoirs) also show similar kind of behaviour for the initial short time dynamics of FVF in the comparative study of Markovian versus non-Markovian cases.\\
\indent
Now, we analyze the time evolution of the FVF for three different types of reservoirs (Ohmic, sub-Ohmic and super-Ohmic) for the both off-resonant and resonant regimes. In general the qualitative behaviour  for all three types of reservoirs are similar in the resonant and off-resonant regimes. Unlike the heating function dynamics, where exchange of energy  between the system and the environment causes non-Markovian oscillation, the FVF dynamics is flat for all the three types of reservoirs. One can also notice that the decoherence process is notable amount faster for resonant case ($r_{1,2}<<1$) than that of off-resonant case ($r_{1,2}>>1$). The system oscillator frequencies lie within the most strongly coupled modes of the reservoir. It results in a strong effective coupling between the system and reservoir. So, one may call the system is on resonance with the reservoir. In addition, the system interacts with the reservoirs through two more modes (counter-rotating terms) for each channel ($1$ and $2$).\\
\indent
 Among all the three type of reservoirs, Ohmic reservoir impels the slowest decoherence. On the other hand, super-Ohmic and sub-Ohmic reservoirs decay in a similar manner, but both decay much faster than the Ohmic reservoir. For time $t << t_{th}^{1,2}$, the exponential factors in Eqs. (24) and (27) can be approximated to one and the FVF dynamics is determined solely by the diffusion coefficient $\Delta_{1,2}(t)$ through the heating function $N_{1,2}(t)$. Thus, one can say that the decoherence rate is highly dependent on the diffusion coefficients $\Delta_{1,2}(t)$. Since, $\Delta_{1,2}(t)$ is highly dependent on the nature of spectrum (through $J(\omega)$), magnetic field, and confining length of the system (through $\omega_{1,2}$ which includes both cyclotron frequency $\omega_c$ and the system oscillator frequency $\omega_0$). The basic message is that the tuning of external magnetic field or the confining length will help us to modify $\Delta_{1,2}(t)$ and results in control of decoherence process. In the next section, we investigate another way to modify the decoherence process based on an elegant approach of Zeno effect, i.e. by performing frequent measurements on the system.
\indent
From the above analysis, it may not be clear whether the observed differences in the FVF dynamics between the Ohmic, super-Ohmic and sub-Ohmic baths is really due to the low frequency behaviour of the bath spectrum. It is more likely that the bath spectral density plays a major role in controlling the FVF dynamics. It seems much more likely that the bath density at resonance frequency is an important quantity and the bath density of course depends on the low frequency behaviour of the environment within the specific choice of the model reservoir (Eq. 18). There is a intuition that the environment induced decoherence is highly dependent on the low frequency behaviour of the environment. This is because if the information of the quantum phases leave the individual system and spread out in the environment without returning back again at low frequencies. In order for this to happen it is not only necessary that the environment is large, but its recurrence time should also be large. As in our case the environment consists of a large number of harmonic oscillators with the same frequency $\omega$. As result of that there would be typically a recurrence time of the order of $\omega^{-1}$. So one can argue that the quantum phases would reappear in the quantum system after this time. One may sense it as similar to the Poincaré recurrence of classical mechanics. It is necessary that the
recurrence time should be large. For our heat bath which consists of large number of harmonic oscillator this is guaranteed by the strong presence of infra-red modes, i.e. $\omega\rightarrow 0$ in the spectral density.\\
\indent
One can notice that the different form of the spectral distribution functions in the Ohmic, sub-Ohmic, and super-Ohmic cases establish a clear connection between the reservoir properties and the dynamics
of Fringe visibility function. The $r_{1,2}=10$ case is known as the off-resonant case where the effective coupling between the system oscillators and the environment is very small for all the three reservoirs mentioned above. This is because the system oscillators are detuned from the peak of the reservoir spectral distribution function. It is observed that the effective coupling between the system and reservoir increases rapidly as $r_{1,2}$ decreases from $10$ to $0.1$ for the Ohmic and sub-Ohmic cases. On the other hand, the super-Ohmic spectral distribution function shows highest coupling at $r_{1,2}=1$ and the cases $r_{1,2}=0.1$ and $r_{1,2}=10$ shows relatively weak coupling. At high temperatures, the sub-Ohmic spectral distribution function has a divergency point at $\omega=0$ which causes very large effective coupling at low frequencies of heat bath spectrum. But, the sub-Ohmic spectral distribution function does not diverge at zero temperature at $\omega=0$. From the expressions of FVF, one can easily notice that FVF depends on heating functions $N_{1,2}$ as well as on the normal dissipation coefficients $\gamma_{1,2}(t)$. At high temperatures, $\Delta_{1,2}>>\Gamma_{1,2}(t)$ which enforces that the FVF dynamics is mostly determined by the behaviour of the heating functions. Heating functions are related with normal diffusion coefficients and the sign of diffusion coefficients determine whether the heating function grows monotonically or oscillatory in nature. It was proved earlier that the non-Markovian heating shows two types of behaviour : an oscillatory behaviour and a monotonic increase \cite{manis1}. It is seen that the oscillations in heating functions for the Ohmic spectrum with $r_{1,2}=10$ originates from the low frequency part of the spectrum. On the other hand, the monotonic increasing part arises due to the resonance part of the spectrum. This connection between the dynamics of heating function and the features of spectrum also hold for other types of reservoirs. The connection between the low frequencies of the heat bath spectrum and oscillations in the heating dynamics can be demonstrated by observing the parameter region where $\Delta_{1,2}(t)$ attains temporarily negative values which gives rise to oscillations in heating functions. For these regime, the peaks of all the three spectral distributions are centred in the low frequencies region. This indicates that the presence of oscillations and the low frequency part of the spectrum are interlinked. The presence of oscillations in heating functions is the indication  of non-Markovian effects for a given value of $r_{1,2}$ and it  depends on the type of reservoir spectrum. The persistence of the memory effects in the sub-Ohmic reservoir is much greater than in the other reservoir types. These features are reflected in the FVF dynamics. The FVF contains heating function in the denominator of the argument of the exponential. Hence the oscillatory behaviour can't be observed and a decay nature of the FVF function is observed as a function of time.
\section{Decoherence control: Quantum Zeno Effect}
   It is well known that the decay of an unstable system can be modified (slowed down) by measuring the system frequent enough \cite{mishra} and is known as quantum Zeno effect (QZE). In some systems an enhancement of decay rate may occur due to frequent measurements, which is known as anti-Zeno effect (AZE) \cite{lane}. Maniscalco {\it et al.} \cite{manis4} demonstrate the occurrence of QZE or AZE for quantum Brownian motion which arises due to the short time non-Markovian behaviour of the environment induced decoherence. Thus, one can use the QZE or AZE to control the quantum-classical crossover by extending or shortening the quantum features of the initial state of the system. In this paper, we extend the study in Ref. \cite{manis4} by incorporating an external magnetic field. Moreover, the important role played by three different reservoirs (Ohmic, sub-Ohmic and super-Ohmic) and the external magnetic field in exploring the crossover between Zeno and anti-Zeno dynamics is investigated in details in the present section. Such study is very important in the sense that we can acquire knowledge on controlling decoherence mechanism for different physical realizations, e.g., a solid state qubit experienced sub-Ohmic noise, while optical qubits interact with flat Markovian spectrum. Unlike, the decoherence control mechanism based on reservoir engineering technique (as described in Sec. 3), different reservoir spectrum lead to different decoherence dynamics in this non-selective energy measurement based Zeno control mechanism.\\
   \indent
 The discussion of a Fock state in the context of quantum information may be a debatable issue as the Fock states are extremely fragile with respect to the environmental coupling. A Fock state with n photons is very hard to generate and very much fragile. In general, a Fock state state with n photons survive for $T_c/n$ (on an average) in a cavity of damping time $T_c$ before undergoing a jump to $|n-1>$ Fock state. Thus, the decoherence rate of such mesoscopic photon number or Fock states is proportional to the photon number. But, recent experiments based on quantum feedback mechanism stabilize Fock states of light in a cavity \cite{haroche1,haroche2,haroche3}. In Refs. \cite{haroche1,haroche3}, the authors demonstrate the recent development of  two quantum feedback schemes which continuously sustaining Fock state of a microwave field in a high-quality cavity. In Ref.\cite{haroche1}, the authors successfully implemented the feedback mechanism to prepare photon number states (Fock states) of a microwave field in a superconducting cavity which can suppresses the effects of decoherence-induced field quantum jumps. They use a beam of atoms crossing the cavity as a sensor and it makes weak quantum non-demolition (QND) measurements of the photon number repeatedly. Between the measurements, a real-time computer (acts as controller) injects adjustable small classical fields in the cavity. A quantum oscillator generates  microwave field and is suitable to act as a quantum memory or as a quantum bus. This active control mechanism which can generate non-classical states of this oscillator and reduces their decoherence. This experiment is a great leap towards the execution of complex quantum information operations. On the other hand, Zhou et. al.\cite{haroche3} creates Fock states with photon numbers (n up to 7) and can be created on requirement in a microwave superconducting cavity by a quantum feedback procedure that reduces decoherence-induced quantum jumps. In this mechanism, circular Rydberg atoms are implemented as QND sensors or as single-photon emitter or absorber actuators. Both schemes efficiently detect the quantum jumps of the photon number and effectively reduce their adverse effects. As a result both mechanisms based on real-time quantum feedback prepares and stabilizes photon number states. These experiments \cite{haroche1,haroche3} create the opportunity to generate stable photon number states which makes the doorway for many opportunities in quantum information. Many applications in quantum information or quantum computing require radiation with a fixed number of photons. This increased the demand for systems able to produce such fields. For example, one can mention the demand of such Fock states for secure quantum communication \cite{a,b} and quantum cryptography \cite{c}. These facts motivate us to discuss the decoherence control in the context of Fock states. \\
   \indent
   The system is initially prepared in the fock state $|n_{1};n_{2}>$. Here, $n_|1>$ denotes the photon number state for $1$ mode with n photons and $n_|2>$ denotes the photon number state for the  mode $2$ . This is just a reminder to the reader that wherever, we use the notation {1,2} as subscript to some variable, it actually implies to specify the $1$ mode and mode $2$ in the combined form. Now, the system is forced to N nonselective energy measurements (i.e. measurements which do not select different outcomes) \cite{breuer} during the time evolution of the system  at a time interval $\tau$ between the two successive measurements.  This $\tau$ is chosen so short that the second order processes may be neglected. This is to be remembered that the meaning of the energy measurement denotes the measurement of the total energy of the system. Since, $\tau$ is small, one can say that the survival probability, i.e., the probability of finding the system in its initial state  $P_{n_{1},n_{2}}(t)=<n_{1},n_{2}|\rho_s(t)|n_{1},n_{2}>$ is such that $P_{n_{1},n_{2}}(t)\simeq 1$ with $P_{n_{1};n_{2}}(t)>>P_{n_{1}+1,n_{2}+1}(t)$. Now, the survival probability after N such successive nonselective measurements is given by \cite{facchi,manis4},
    \begin{equation}
    P^{N}_{n_{1},n_{2}}(t)=\lbrack P_{n_{1};n_{2}}(\tau)\rbrack^N\equiv exp(-\gamma_{n_{1,2}}^z(\tau),
    t)
    \end{equation}
   \indent
    where $\gamma_{n_{1,2}}^z(\tau)$ is the effective decay rate and t=$\tau N$ is the effective duration of the experiment. Here,  $\gamma_{n_{1}}$  denotes the decay rate for the $1$ mode and $\gamma_{n_{2}}$  for the  mode 2. We write them in a combined form as : $\gamma_{n_{1,2}}$.  The effective decay rate, which is defined in Eq. (30), determines the occurrence of the Zeno or anti-Zeno effect . Let us define $\gamma_{n_{1,2}}^0$ as the decay rate of survival probability in the absence of measurement. This corresponds to the Markovian decay rate in the limit $\tau\rightarrow\infty$ :
    \begin{equation}
    \gamma_{n_{1,2}}^0 = \lim_{\tau\rightarrow\infty}\gamma_{n_{1,2}}^Z(\tau)=\Delta_{M},
    \end{equation}
    where $\Delta_{M}=\pi I(\omega)$ with $I(\omega)=J(\omega)\lbrack N(\omega)+1/2\rbrack$. Thus, it is evident that $\gamma_{n_{1,2}}^z(\tau)$ depends on the reservoir spectrum, external magnetic field and the oscillator frequency. If there exist a finite time $\tau^*$ in such a way that $\gamma_{n_{1,2}}^z(\tau^*)=\gamma_{n_{1,2}}^0$, then one can say that $\frac{\gamma_{n_{1,2}}^z(\tau)}{\gamma_{n_{1,2}}^0}<1$ for $\tau<\tau^*$. This implies that the measurements slow down the decay and we are at QZE regime. On the other hand, one can find $\frac{\gamma_{n_{1,2}}^z(\tau)}{\gamma_{n_{1,2}}^0}>1$ for $\tau>\tau^*$ which accelerates the decay of the system (AZE) \cite{16}.
    \indent
    Let us consider that our initial states are Fock-Darwin states $|n_{1},n_{2}>$. Each of these initial states are associated with the decay channels corresponding to the upward transitions  $|n_{1}+1;n_{2}+1>$ and downward transitions  $|n_{1}-1;n_{2}-1>$,  respectively. Now, following the method introduced in Ref. \cite{manis4}, one can show that the effective decay rate is given by :
    \begin{eqnarray}
    \gamma_{n_{1,2}}^Z(\tau)&=&\frac{1}{\tau}\big[(2n_{1,2}+1){\int_{0}^{\tau}}\Delta_{1,2}(t)dt \nonumber \\&-&{\int_{0}^{\tau}}\Gamma_{1,2}(t)dt\big]
    \end{eqnarray}
    where, $\Delta_{1,2}(t)$ and $\Gamma_{1,2}(t)$ are given in equations (14) and (15) respectively.
    The Markovian value of effective decay rate in the limit $\tau\longrightarrow\infty$,
    \begin{equation}
    \gamma_{n_{1,2}}^0=(2n_{1,2}+1)\Delta_{M} - \Gamma_{M}
    \end{equation}
    Where $\Delta_{M}$ and $\Gamma_{M}$ are the Markovian values of diffusion and dissipation coefficients respectively. The quantity which governs the crossover between the Zeno and anti-Zeno dynamics is the ratio:
    \begin{equation}
    \frac{\gamma_{n_{1,2}}^Z(\tau)}{\gamma_{n_{1,2}}^0}=\frac{(2n_{1,2}+1)\int_{0}^{\tau}\Delta_{1,2}(t)dt-\int_{0}^{\tau}\Gamma_{1,2}(t)dt}{\tau(2n_{1,2}+1)\Delta_{M} -\Gamma_{M}}.
    \end{equation}
    This is well understood that the above mentioned ratio is highly dependent on the initial Fock-Darwin states $|n_{1},n_{2}>$. At high temperatures, the diffusion term $\Delta_{1,2}(t)$ dominates over the dissipative terms $\Gamma_{1,2}(t)$,i.e., $\Delta_{1,2}(t) >> \Gamma_{1,2}(t)$ and we obtain
    \begin{equation}
    \frac{\gamma_{n_{1,2}}^Z(\tau)}{\gamma_{n_{1,2}}^0}=\frac{{\int_{0}^{\tau}}\Delta_{1,2}(t)dt}{\tau\Delta_{M}},
    \end{equation}
and the ratio is independent of initial state. These two equations (32) and (33) make a connection between two fundamental quantum phenomena, e.g., the quantum Zeno effect (QZE) and environment induced decoherence (EID). In the rest of this section, we elucidate this connection between QZE and EID and discuss its physical relevances. It is to be mentioned that Eqs. (32) and (33) are derived without any assumption on the form of spectral density of the reservoir.  Now we analyze these two Eqs. (32) and (33) to observe Zeno and anti-Zeno crossover and the control mechanism of this crossover.\\
    \fignine
    \figten
    \indent
    Now, one can say that the effective decay rate at high temperatures solely depend on the diffusion coefficients $\Delta_{1,2}(t)$ . Hence, the crossover from QZE and AZE and their relation with EID mainly depends on the behaviour of $\Delta_{1,2}(t)$ at high T. This can be explained as follows. The average EID, (between two successive measurements) which is quantified by $\frac{\int_{0}^{\tau}\Delta_{1,2}(t)dt}{\tau}$, is less than the Markovian case (quantified by $\Delta_{M}$) then the effect of EID is less than the Markovian one. As a result of the measurements the system experiences repeatedly an effective EID which is less than the Markovian one. So, one can observe QZE. On the other hand, if the average EID measured between two succesive measurements is greater than the Markovian value $\Delta_{M}$, the system feels strong decoherence which accelerates the decay and we experience AZE. Thus, one can infer an important conclusion that the quantum-classical transition can be controlled by either QZE or by AZE. As we know that the energy eigenstates (Fock-Darwin States) of the harmonic oscillator are highly delocalized, one can prolonged an initial delocalized energy eigenstate by means of QZE and vice versa by AZE.\\
    \indent
    One can find the relevant parameters, which controls the crossover between the QZE and AZE, by observing the expression of diffusion coefficients $\Delta_{1,2}(t)$ (Eq. 14). From this expression, it is understood that the transition is governed by the ratio $r_{1,2}=\frac{\omega_{1,2}}{\omega_{cut}}$ which quantifies the asymmetry of the spectral distribution and the ratio $\frac{k_BT}{\hbar\omega_{1,2}}$. In this respect we need to introduce two more parameters $r_c=\frac{\omega_c}{\omega_{cut}}$ (related to external magnetic field through cyclotron frequency $\omega_c=\frac{eB}{mc}$) and $r_0= \frac{\omega_0}{\omega_{cut}}$ (related to confinement length through $\omega_0$). Thus, one can rewrite $r_{1,2}=\sqrt{r_0^2+r_c^2/4}\pm r_c/2$. Hence, one can transit from Zeno regime to anti-Zeno regime and vice versa by changing the values of the parameters $r_c$, $r_0$ and $T$. As a result of that one can control EID by tuning the parameters $r_c$, $r_0$ and $T$. On the other hand, the picture changes dramatically at zero temperature which is recognized by the asymmetric spectral density function. As Eq. (32) governs the dynamics, the QZE-AZE crossover exists even at the edge of the spectral density function where $\omega_{1,2}\simeq \omega_{cut}$. This is mainly because of the presence of initial jolt in the diffusion coefficients $\Delta_{1,2}(t)$ and in the dissipation coefficients $\Gamma_{1,2}(t)$ and this is the signature of strong decoherence. \\
    \indent
    We plot the ratio of the effective decay rate $\gamma_{n_{1,2}}^z(\tau)$ and the Markovian decay rate $\gamma_{n_{1,2}}^0$ as a function of dimensionless time for three types of reservoirs Ohmic, sub-Ohmic and super-Ohmic in Figs. (3-6) with different relevant parameters. The sign $1,2$ denotes two different channels of the system. Figures (3-6) represent the decay behaviour for $r_{1,2}>>1$ regime at high temperatures (Figs. 3 and 4) as well as at zero temperature (Figs. 5 and 6). On the other hand, figures (7-10) demonstrate the dynamical behavior of the ratio between the effective decay rate to that of the Markovian one in the regime with $r_{1,2}<<1$ for high temperatures (Figs. 7 and 8) in addition to zero T behaviour (Figs. 9 and 10). In the regime $r_{1,2}>>1$, we observe that QZE predominates for Ohmic (uppermost row) and super-Ohmic (lowest row) reservoirs at high temperatures as shown in Fig. (3) and Fig. (4) with a variation of $r_c$ and $r_0$ respectively. On the other hand, sub-Ohmic reservoir (middle row in Figs 3 and 4) shows a cross-over  between Zeno and anti-Zeno dynamics (mostly the $1$ channel) either by varying magnetic field ($r_c$) or by the variation of confinement length ($r_0$). Let us move to the analysis of the case at zero-T for $r_{1,2}>>1$ regime (Figs 5 and 6). In this regime, both the Ohmic (uppermost row) and sub-Ohmic (middle row) reservoirs show QZE to AZE crossover. This can be achieved either by varying magnetic field $r_c$ (Fig. 5) or by the variation of length of confinement $r_0$ (Fig. 6). In both the cases, we observe that the dynamics is dominated by QZE for the super-Ohmic reservoir (lowest rows in Figs. 5 and 6).\\
    \indent
    It is always found that the method of nonselective energy measurements accelerate EID in the $r_{1,2}>>1$ regime. The initial shake in the diffusion coefficients $\Delta_{1,2}(t)$ and it makes an initial decoherence much powerful than the Markovian one \cite{manis4}. This off-resonant regime is denoted by vigorous non-Markovian hallmark such as oscillations.
    \indent
    Now, we move to the discussion of the dynamic behaviour of the ratio $\frac{\gamma_{n_{1,2}}^z(\tau)}{\gamma_{n_{1,2}}^0}$ for the regime with $r_{1,2}<<1$ at high temperatures (Figs. 7 and 8) as well as at zero T (Figs. 9 and 10). We always observe QZE for all the three type of reservoirs mentioned above at high temperatures for the variation of $r_c$  (Fig. 7) as well as for the variation of $r_0$ (Fig. 8). But the scenario is quite different for the super-Ohmic cases in Fig. (7) and Fig. (8). For the super-Ohmic reservoir at high temperatures, one can observe Zeno to anti-Zeno crossover for relatively high magnetic field $r_c=0.5$  with constant $r_0=0.5$ (see rightmost column and lowermost row in Fig. 7) and the same kind of crossover can be found by varying  the confinement length $r_0$ at a constant $r_c=0.05$ (see the lowermost row of Fig. 8). Unlike the $r_{1,2}>>1$ regime, where the crossover between Zeno and anti-Zeno is observed in the $1$ channel, the crossover  is mainly found in the $2$ channel. Let us discuss the zero-T behaviour. At zero-T, one can observe the Zeno to anti-Zeno crossover for all the three type of reservoirs mentioned above. Again this crossover is mainly observed for the $2$ channel. One can summarize that changing three parameters i.e. temperature ($T$), external magnetic field $r_c$ and the confinement length $r_0$ one can pass from QZE to AZE and hence one can control EID. \\
    \indent
    In Fig. 11, we display the contour plot of the effective decay rate ratio at high temperatures (Eq. 33) for three different reservoirs ( Ohmic, sub-Ohmic and super-Ohmic) as a function of the control parameter $r_{1,2}$ (involves both $r_0$ and $r_c$) and of the measurement interval $\omega_{1,2}\tau$. The QZE to AZE crossover is mainly specified by some bold solid lines. It is observed that QZE dominated region, i.e. $\tau^*$ does not exist, for some values of $r_{1,2}$ for the Ohmic as well as super-Ohmic reservoirs. Also, one can find some region  for the  super-Ohmic  spectra which corresponds two $\tau^*$. This represents that there are two $\tau^*$ which correspond to the crossover from Zeno to anti-Zeno and then again to Zeno dynamics as one increase $\omega_{1,2}\tau$. The Ohmic reservoir shows that the AZE occurs for $r_{1,2}<<1$. On the other hand, one can always found Zeno to anti-Zeno dynamics, i.e., a scenario in which the EID  decelerates to a circumstances where it accelerates. The contour plot informs us about the probable values of $r_c$, $r_0$ and measurement interval $\tau$ for which we can decelerate decoherence and vice versa.\\
    \indent
  The effective decay rate extensively depends on the overlap of the spectral density of the heat bath  and the system oscillator frequencies. Thus, one can expect that changing the environment spectral density one can easily control the decay rate, at least quantitatively. The  Zeno to anti-Zeno transitions observed in the contour plot can be described as follows. Mostly, in the resonant regime $r_{1,2}<<1$, the frequency of the oscillator overlaps with the reservoir spectrum. As a result of that measurements performed at times smaller
than the reservoir correlation time $\tau_R$ strongly inhibit the quantum to classical transition (QZE). On the other hand, in the off-resonant regime  the anti-Zeno effect prevails. The effective coupling between the system and the reservoir for the Ohmic and sub-Ohmic reservoirs becomes stronger when $r_{1,2}$ grows from 0.1 to 10. On the other hand, the super-Ohmic reservoir shows the highest effective coupling for $r_{1,2} = 1$, and the cases $r_{1,2} = 0.1$ and $r_{1,2} = 10$ correspond to relatively weak couplings, i.e., off-resonant regimes. Henceforth, we observe two AZE regimes separated by a QZE regime for the super-Ohmic case.
    \figeleven
    \subsection{Probable Experiment}
    The kind of study conducted here can be realized experimentally in the context of trapped ion. The recent
advancement in the field of laser cooling and trapping techniques make it possible to confine a single ion in a harmonic well at very low temperatures. Hence, one can use a miniature version of the linear Paul trap to trap the ion \cite{raizen,jefferts}. A single laser cooled ion is theoretically equivalent to a charged
particle moving in a harmonic well. On the other hand, the recent development in reservoir engineering techniques paves the way for constructing different type of reservoirs as  considered in this paper. Not only it is possible to construct an "artificial" reservoir but also one can manipulate its spectral density and the coupling with the system \cite{turchette}. The possible way to implement our QBM model for an Ohmic and a sub-Ohmic environment has been discussed in Refs. \cite{rajesh}. The same technique can be adopted straightforwardly to realize the Ohmic and sub-Ohmic environments for a trapped ion. The Ohmic and sub-Ohmic environment can be modeled by an infinite RLC
transmission line. The transmission line is made of discrete building blocks consist of inductor (L) and resistor (R) in series along one stringboard of the ladder and the capacitor (C) is on the horizontal support of the ladder. Thus it is evident that the Ohmic and sub-Ohmic environment can be realized from the LC-dominant and R-dominant limit of the RLC transmission line, respectively \cite{rajesh}. On the other hand, the nonselective energy measurement technique as discussed in this paper can be realized experimentally by the filtering process of a random electric field which is coupled to a trapped ion by its charge q, $H_{SB} = −q\vec{r}\cdot\vec{E}$ , and is equivalent to the bilinear coupling as discussed in this paper. Now, controlling the switch off-on of this noisy electric field could be thought of as physical implementation of the nonselective energy measurements. The sudden switch off-on of this noisy electric field can be thought of physical realization of the operation of trace over the reservoir degrees of freedom. As matter of fact one can consider this trace operation as a physical implementation of nonselective energy measurements (see Ref. \cite{1}, page 321). Thus, controlling the switch off-on procedure (i.e. measurement interval $\tau$) of the noisy electric field by the shuttering process one can pass from Zeno to anti-Zeno regime depending on the values of the system parameters (e.g. $\omega_0$), reservoir parameters (e.g.,$T$ and $\omega_{cut}$), and external field ($\omega_c$). The required values can easily be obtained from the contour plot (Fig. 11). All these parameters ($\omega_0$, $\omega_c$, $\tau$, $T$ and $\omega_{cut}$), which drives the Zeno to anti-Zeno crossover, can easily be tuned in this proposed experiment. It is to be mentioned here that the parameter $\omega_0$ can be modified within a short range, but the modifications of $\omega_c$, $\omega_{cut}$, and $T$ can be acquired by merely varying external magnetic field, filtering the applied noisy electric field and changing the noise fluctuations respectively \cite{6}. Since, all the parameters controlling the crossover between Zeno to anti-Zeno can  easily be tuned, the observation of such crossover and the control of EID in the context of trapped ion can be in the grip of experimentalists.
\section{Discussion}
In the present paper, we analyze two different techniques for manipulating environment induced decoherence of a charged quantum oscillator in contact with three different types of Bosonic environments in the presence of an external magnetic field. First, we discuss the strategy based on the reservoir engineering technique to manipulate the quantumness of an initial Schr$\ddot{o}$dinger cat state. The main finding is that Ohmic reservoir is more suitable than the sub-Ohmic or super-Ohmic reservoir to sustain the quantumness of the cat state. The transformation from the initial non-classical cat state to the classical statistical mixture is much slower for the Ohmic case than that for the sub-Ohmic or super-Ohmic reservoirs. Therefore, if one can tune the reservoir spectrum into an Ohmic form, it is possible to slow down the decoherence processes. It is to be mentioned here that the quantum-classical transition is characterized by the disappearance of the interference pattern of the corresponding Wigner function. We investigate the short time non-Markovian dynamics of the dissipative charged magneto-oscillator. One can obtain the solutions in terms of Wigner function for the two regimes,i.e. for the off-resonant ($r_{1,2}>>1$) and resonant regime ($r_{1,2}<<1$). The solutions essentially tell us that the counter rotating terms has significant contribution in the resonant regime, while it has negligible effect in the off-resonant regime. The analysis of our solutions show us that the environment induced decoherence (EID) strongly depends on the diffusion coefficients $\Delta_{1,2}(t)$ through the heating functions ${N}_{1,2}(t)$. Although, both the decoherence and the heating or dissipation processes are characterized by the same coefficients $\Delta_{1,2}(t)$, the time scales of these two processes are distinctly different for our system in the regime $t<<t_{th}^{1,2}$. As, $\Delta_{1,2}(t)$ are highly sensitive to the system or reservoir parameters, e.g., magnetic field $r_c$, confinement length $r_0$ or the cut-off frequency $\omega_{cut}$, one can modify the quantum-classical transition of the cat-state by tuning the above mentioned parameters. Also, one can observe that the heating caused by the Ohmic reservoir is much slower than that of the super-Ohmic or sub-Ohmic cases. Hence, one can modify reservoir spectrum to Ohmic type to observe slowest decoherence.\\
\indent
Now, the second procedure for controlling EID is based on quantum Zeno effect (QZE). This investigation tell us that the quantum-classical transition of an initial Fock-Darwin state of the quantum charged oscillator is highly sensitive to the reservoir spectrum and the parameters $r_c$, $r_0$, $\tau$, and $\omega_{cut}$. We propose a possible experimental realization in the context of trapped ion for modifying Zeno-anti-Zeno crossover by tuning the above mentioned parameters. The controlled continuous moving from QZE (decay supression) to AZE (decay acceleration) can be manipulated either by tuning above mentioned parameters or varying the reservoir spectrum. Thus, the second technique is easier to implement in experiment to move in a controlled continuous passage from quantum to classical state via QZE and AZE.
\begin{acknowledgments}
MB acknowledge the financial support of IIT Bhubaneswar through seed money project SP0045. AMJ thanks DST, India  for award of J C Bose national fellowship.
\end{acknowledgments}
	

\begin{thebibliography}{99}
  	\bibitem{1}H.-P. Breuer and F. Petruccione, {\em The Theory of Open Quantum Systems} (Oxford University Press, Oxford, 2002).
  	\bibitem{2}U. Weiss, {\em Quantum Dissipative Systems} (World Scientific, Singapore, 2008).
  	\bibitem{3}W. H. Zurek, {\em in Physical Origins of Time Asymmetry} (Cambridge University Press, Cambridge, 1994); W. H. Zurek, Rev. Mod. Phys. {\bf 75}, 715 (2003).
  \bibitem{unruh}W.G. Unruh, Phys. Rev.A {\bf 51}, 992 (1995). See also I.L. Chuang, R. Laflamme, P.W. Shor,
W.H. Zurek, Science, {\bf 270}, 1633 (1995); W.H. Zurek and J.P. Paz, Il Nuovo Cimento, {\bf 110B},
611 (1995).
  	\bibitem{4}W. T. Strunz, F. Haake, and D. Braun, Phys. Rev. A {\bf 67}, 022101 (2003); M. S. Kim and V. Buzek, Phys. Rev. A {\bf 46}, 4239 (1992).
  	\bibitem{5}J. Paavola, J. Piilo, K.-A. Suominen, and S. Maniscalco, Phys. Rev. A {\bf 79}, 052120 (2009); J. Paavola, and S. Maniscalco, Phys. Rev. A {\em 82}, 012114 (2010).
  	\bibitem{6}S. Maniscalco, J. Piilo, F. Intravaia, F. Petruccione, and A. Messina, Phys. Rev. A {\bf 69}, 052101 (2004); ibid {\bf 70}, 032113 (2004).
    \bibitem{viola}G. M. Palma, K. A. Suominen, and A. K. Ekert, Proc. R. Soc. London, Ser. A {\bf 452}, 567 (1996); L. M. Duan and G. C. Guo, Phys. Rev. Lett. {\bf 79}, 1953 (1997); P. Zanardi and M. Rasetti, Phys. Rev. Lett. {\bf 79}, 3306 (1997); D. A. Lidar, I. L. Chuang, and K. B. Whaley, Phys. Rev. Lett. {\bf 81}, 2594 (1998); L. Viola and S. Lloyd, Phys. Rev. A {\bf 58}, 2733 (1998); D. Vitali and P. Tombesi, Phys. Rev. A {\bf 59}, 4178 (1999); L. Viola, E. Knill, and S. Lloyd, Phys. Rev. Lett. {\bf 85}, 3520 (2000); A. G. Kofman and G. Kurizki, Phys. Rev. Lett. {\bf 87}, 270405 (2001); M.S. Byrd and D. A. Lidar, Phys. Rev. A {\bf 67}, 012324 (2003).
        \bibitem{facchi}P. Facchi, S. Tasaki, S. Pascazio, H. Nakazato, A. Tokuse, and D. A. Lidar, Phys. Rev. A {\em 71}, 022302 (2005).
            \bibitem{manis1}S. Maniscalco, Laser Physics {\em 20}, 1251 (2010)
            \bibitem{manis2}J. Paavola and S. Maniscalco, Phys. Rev. A {\em 82}, 012114 (2010)
            \bibitem{ford} G. W. Ford, J. T. Lewis, and R. F. O'Connell, Phys. Rev. A {\em 37},
             4419 (1988).
             \bibitem{amico}L. Amico, R. Fazio, A. Osterloh, and V. Vedral,Rev. Mod. Phys. {\em 80}, 517 (2008)
             \bibitem{srednicki}M. Srednicki, Phys. Rev. Lett. {\em 71}, 666 (1993); K. Audenaert, J. Eisert, M.  B. Plenio, and R. F. Werner, Phys. Rev. A {\em 66}, 042327 (2002); M. B. Plenio, J. Hartley, and J. Eisert, New. J. Phys. {\em 6}, 36 (2004).
                 \bibitem{eisert}J. Eisert, M. B. Plenio, S. Bose, J. Hartley, Phys. Rev. Lett. {\em 93}, 190402 (2004)
                     \bibitem{galve1}F. Galve, and E. Lutz, Phys. Rev. A {79}, 032327 (2009)
                     \bibitem{galve2}F. Galve, G. L. Giorgi, and R. Zambrini, Phys. Rev. A {\em 81}, 062117 (2010)
             \bibitem{hu}B. L. Hu, J. P. Paz, and Y. Zhang, Phys. Rev. D {\em 45}, 2843 (1992).
  	\bibitem{7}F. Intravaia, S. Maniscalco, and A. Messina, Phys. Rev. A {\bf 67}, 042108 (2003); ibid Eur. Phys. J. B {\em 32}, 97 (2003).
  \bibitem{manis3}S. Maniscalco, J. Piilo, and K. -A. Suominen, Eur. Phys. J. D {\em 55}, 181 (2009)
  \bibitem{kim}M. S. Kim and V. Buzek, Phys. Rev. A {\em 46}, 4239 (1992)
  \bibitem{myatt}C. J. Myatt, B. E. King, Q. A. Turchette, C. A. Sackett, D. Kielpinski, W. M. Itano, C. Monroe, and D. J. Wineland, Nature {\em 403}, 269 (2000).
      \bibitem{mishra}B. Mishra and E. C. G. Sudarshan, J. Math. Phys. (NY) {\em 18}, 756 (1977).
      \bibitem{manis4} S. Maniscalco, J. Piilo, and K-A. Suominen, Phys. Rev. Lett. {\em 97}, 130402 (2006)
      \bibitem{lane}A. M. Lane, Phys. Lett. A {\em 99}, 359 (1983); A. G. Kofman and G. Kurizki, Nature (London), {\em 405}, 546 (2000).
          \bibitem{breuer}H.-P. Breuer and F. Petruccione, {\it The Theory of Open Quantum Systems} (Oxford University Press, New York, 2002)
              \bibitem{haroche1}C. Sayrin, I. Dotsenko, X. Zhou, B. Peaudecerf, T. Rybarczyk, S. Gleyzes, P. Rouchon, M. Mirrahimi, H. Amini, M. Brune, Jean-Michel Raimond, S. Haroche, Nature {\em 477}, 73 (2011)
                  \bibitem{haroche2}B. Peaudecerf, C. Sayrin, X. Zhou, T. Rybarczyk, S. Gleyzes, I. Dotsenko, J. M. Raimond, M. Brune, and S. Haroche, Phys. Rev. A {\em 87}, 042320 (2013)
                      \bibitem{haroche3}X. Zhou, I. Dotsenko, B. Peaudecerf, T. Rybarczyk, C. Sayrin, S. Gleyzes, J. M. Raimond, M. Brune, and S. Haroche, Phys. Rev. Lett. {\em 108}, 243602 (2012)
                          \bibitem{a} H. Zbinden et al., J. Cryptol. {\em 13}, 207 (2000); H.-K. Lo and
H. F. Chau, Science {\em 283}, 2050 (1999).
\bibitem{b} K. M. Gheri et al., Phys. Rev. A {\em 58}, R2627 (1998); S. J.
van Enk et al., Phys. Rev. Lett. {\em 78}, 4293 (1997); S. J.van Enk et al., Science {\em 279}, 205 (1998).
\bibitem{c}T. Jennewein {\it et al.}, Phys. Rev. Lett. {\em 84}, 4729 (2000); D. S.
Naik {\it et al.}, Phys. Rev. Lett. {\em 84}, 4733 (2000); W. Tittel {\it et al.}, Phys. Rev. Lett. {\em 84}, 4737 (2000).
\bibitem{15}P. Facchi and S. Pascazio, J. Phys. A : Math. Theor. {\em 41}, 493001 (2008)
              \bibitem{16}A. Shaji, J. Phys. A. : Math. Gen. {\em 37}, 11285 (2004)
          \bibitem{raizen}M. G. Raizen, J. M. Gilligan, J. C. Bergquist, W. M. Itano, and
D. J. Wineland, Phys. Rev. A {\em 45}, 6493 (1992).
\bibitem{jefferts}S. R. Jefferts, C. Monroe, E.W. Bell, and D. J.Wineland, Phys.
Rev. A {\em 51}, 3112 (1995).
\bibitem{turchette}Q. A. Turchette, C. J. Myatt, B. E. King, C. A.
Sackett, D. Kielpinski, W. M. Itano, C. Monroe,
and D. J. Wineland, Phys. Rev. A {\em 62}, 053807
(2000).
\bibitem{rajesh}A. Rajesh and M. Bandyopadhyay, Phys. Rev. E {\em 89}, 062116
(2014); ibid Phys. Rev. A {\em 92}, 012105 (2015).
  	\end{thebibliography}
\end{document}